\documentclass[usenatbib]{mnras}

\usepackage{newtxtext,newtxmath}
\usepackage[T1]{fontenc}

\DeclareRobustCommand{\VAN}[3]{#2}
\let\VANthebibliography\thebibliography
\def\thebibliography{\DeclareRobustCommand{\VAN}[3]{##3}\VANthebibliography}

\usepackage{graphicx}	% Including figure files
\usepackage{amsmath}	% Advanced maths commands

\newcommand{\kms}{\,\mathrm{km\,s^{-1}}}	%\,km\,s$^{-1}$, used in Section 4 and Appendix D
\newcommand{\Jzu}{\,\mathrm{kpc\,km\,s^{-1}}}	% units of J_z, used in Section 4 and Appendix D
\newcommand{\nn}{\nonumber}
\newcommand{\cmark}[1]{\textcolor{red}{#1}}	% red: text that is new since the version of 28 Nov 2026, morning
\title[Formation and evolution of the \textit{Gaia} wide binaries]{Formation and evolution of the \textit{Gaia} wide binaries}

\author[L. Thulasidharan et al.]{
Lekshmi Thulasidharan,$^{1}$\thanks{E-mail: lekshmi.t@princeton.edu}
Chris Hamilton,$^{1}$
Nadia Zakamska $^{2,3}$
and Hagai B. Perets$^{4}$
\\  
$^{1}$Department of Astrophysical Sciences, Princeton University, 4 Ivy Lane, Princeton, NJ 08544, USA\\
$^{2}$ William H.\ Miller III Department of Physics and Astronomy, Bloomberg Center, Johns Hopkins University,
3400 N. Charles St., Baltimore, MD 21218, USA\\
$^{3}$ Institute for Advanced Study, Einstein Dr., Princeton, NJ 08540, USA\\
$^{4}$Technion -- Israel Institute of Technology, Technion City, Haifa, Israel, 3200002
}

\pubyear{\the\year{}}

\begin{document}
\label{firstpage}
\pagerange{\pageref{firstpage}--\pageref{lastpage}}
\maketitle

% Abstract of the paper
\begin{abstract}
The eccentricities of wide binary stars carry information about how these binaries formed, but
encounters with passing stars and Galactic tidal torques also evolve them over billions of years.
To separate these two effects, we measured the eccentricity distributions $p(e)\propto e^{\alpha}$
of \textit{Gaia} 
wide binaries as functions of their vertical action $J_z$ in the Galaxy (which increases on average with age) 
and the magnitude contrast $\Delta G$ between their components (which traces their mass
ratio). 
We focused on the $\sim$12,000 binaries with projected separations of $10^3$--$10^{3.5}$\,au within
200\,pc.  We find that at large $J_z$, binary eccentricities are consistent with the 
thermal distribution  $\alpha \simeq 1$ regardless of $\Delta G$. 
However, at low $J_z$, the eccentricities are much more sensitive to $\Delta G$,  
roughly satisfying $\alpha \simeq 2.9 - 0.4 \Delta G$, so that 
binaries with the smallest (largest) $\Delta G$ have very superthermal (subthermal) eccentricities.
We find the same trends when we replace $J_z$ by ages inferred from the stellar parameters of the binaries.
We fit these data by forward-modelling the dynamical evolution of 
wide binaries in the Galactic disc. Our models suggest that 
wide binaries must be formed with eccentricities that depend strongly on their mass ratio, but also, puzzlingly, 
that naive dynamical evolution is too slow to fully explain 
the measured trend of $\alpha$ versus $J_z$. We can come close to reproducing the observations if encounters were much more efficient for the first few Gyr of each binary's life, or if binaries born in the last Gyr 
had different birth eccentricities, but we find neither explanation satisfactory.
\end{abstract}

% Select between one and six entries from the list of approved keywords. 
% Don't make up new ones.
\begin{keywords}
binaries: general -- binaries: visual -- stars: formation -- stars: kinematics and dynamics --
Galaxy: disc
\end{keywords}

%%%%%%%%%%%%%%%%%%%%%%%%%%%%%%%%%%%%%%%%%%%%%%%%%%

%%%%%%%%%%%%%%%%% BODY OF PAPER %%%%%%%%%%%%%%%%%%

\section{Introduction}

Wide binaries are gravitationally bound pairs of stars
 with orbital separations $\gtrsim 10^3$\,au. 
A bound binary follows an elliptical orbit, characterised by its semimajor axis $a$ 
and eccentricity $e$. 
The distributions of these orbital elements likely 
contain information about formation mechanisms 
 that are not yet fully understood \citep{hwang2022,Hamilton2024,mathew2024wide}. 
Extremely eccentric wide binaries and triples have also been proposed as a source of stellar collisions
\citep{Kaib2014,Grishin2022}, of low-mass X-ray binaries \citep{Michaely2016} and of mergers of binary black holes
\citep{Michaely2019}.

In particular, wide binary eccentricity distributions are often well fit (both in data and dynamical models)
 by the power law ansatz
\begin{equation}
p(e)=(1+\alpha)e^\alpha,
\end{equation}
where $\alpha=1$ denotes the so-called \textit{thermal} distribution,
$\alpha>1$ a superthermal distribution, 
and $\alpha<1$ a subthermal distribution \citep{hwang2022}. 
Compared to a thermal distribution, 
a superthermal distribution has a higher fraction of highly eccentric binaries, 
whereas a subthermal distribution has a relatively lower fraction. 
\citet{hwang2022} measured $\alpha$ for \textit{Gaia} binaries in bins of projected separation $s$.
They found that $\alpha$ rises with $s$.
The eccentricity distribution is close to uniform ($\alpha\simeq0$) for $s\lesssim100$\,au, close to thermal
at $s\simeq500$\,au and superthermal beyond $10^3$\,au.
For $10^3<s<10^4$\,au they measured $\alpha\simeq1.3$.
Moreover, the eccentricity distribution seems to depend not just on orbital separation but also on binary
mass ratio $q \equiv m_2/m_1\leq1$.
\citet{hwang2022b} selected \textit{twin} binaries, with $q\gtrsim0.95$, and found that at separations of
100--3000\,au the twins are significantly more superthermal than the other binaries.
%At $400<s<1000$\,au, for example, they measured $\alpha\simeq2.1$ for twins and $\alpha\simeq1.05$ for the
%other binaries.
%Among binaries with $q\lesssim0.9$, however, they found no strong dependence of $\alpha$ on $q$ at any
%separation between 100 and 3000\,au.
%Their samples combined binaries of all ages.

Different formation channels may imprint different eccentricity distributions at birth.
For instance, wide binaries that form when stars in the turbulent interstellar medium pair at random may be 
born with a highly superthermal distribution \citep{Xu2023}.
Wide binaries can also form, or become permanently bound, as a star cluster disperses
\citep{Kouwenhoven2010,Moeckel2011,Rozner2023}, and in the simulations of \citet{Kouwenhoven2010} they are born
with a thermal distribution.
The same process captures free-floating planets on to very wide orbits, so it extends to extreme mass ratios
\citep{Perets2012}.
Wide binaries can also form in a star cluster when three unbound stars meet and two of them end up bound.
This three-body channel mostly produces wide binaries with superthermal eccentricities
\citep{Stone2019,GinatPerets2024,Atallah2024,Atallah2025}.
Formation channels may also differ in how they depend on the masses of the two stars.
Twins are thought to form at separations $\lesssim100$\,au, through accretion from a circumbinary disc
that drives $q$ towards unity, and to be widened later by interactions in their birth cluster
\citep{ElBadry2019twins}.
In simulations of star cluster formation, \citet{mathew2024wide} found that wide binaries with a low total
mass are much more superthermal than those with a high total mass.
In cluster dissolution, by contrast, the mass ratios are close to those expected when stars pair at
random \citep{Kouwenhoven2010}.
Thus, we might hope that by measuring the orbital properties of wide binaries in concert with their mass ratio,
we can infer their dominant
formation processes.

However, the distribution of binary orbital elements observed today may differ
from that at formation,  because those elements evolve over long timescales due to
 perturbations from passing stars and the Galactic tidal torque.
 \cite{Hamilton22,Modak2023,Hamilton2024} studied how these perturbations affect
 wide binary semimajor axis and eccentricity distributions. Repeated gravitational encounters with passing stars produce stochastic changes in the orbital energy and angular momentum of a binary, leading to diffusion in semimajor axis and eccentricity space \citep{bahcall1985maximum,Weinberg1987,caputo2026evolution}.
Galactic tides produce secular changes in orbital angular momentum and hence also change the eccentricity, but on average do not change the semimajor axis. Both processes tend to drive the eccentricity distribution towards thermal (see also \citealt{Nagpal2026}). \cite{Hamilton2024} therefore argued that since the wide binaries with separations $>10^3$\,au are observed to have a  superthermal eccentricity distribution, their birth eccentricity distribution must be \textit{even more} superthermal, and predicted that the eccentricity distribution should become progressively \textit{less} superthermal with increasing binary age $\tau$, i.e., ${d\alpha}/{d\tau}<0$. The rate of eccentricity evolution depends on the semimajor axis of the binary; roughly, thermalization ($\alpha\to 1$) occurs on a characteristic timescale comparable to the disruption timescale of a wide binary \citep[][equation 15]{Hamilton2024}, namely \begin{align}
    \label{eq:ionize}
    t_\mathrm{dis} & \simeq \frac{0.02\sigma m_\mathrm{b}}{G m_\mathrm{p}^2 n a \ln \Lambda}
     \\
    & \simeq 3.8 \, \mathrm{Gyr}\, \left(\frac{\ln \Lambda}{10} \right)^{-1} \left( \frac{m_\mathrm{b}}{M_\odot}\right)
    \left( \frac{m_\mathrm{p}}{M_\odot}\right)^{-2} \left( \frac{n}{0.1 / \mathrm{pc}^{3}}\right)^{-1} \nn \\ & \,\,\,\,\,\,\,\,\,\,\,\,\,\,\,\,\,\,\,\,\,\,\,\times \left( \frac{\sigma}{40\kms}\right) \left( \frac{a}{10^4\,\mathrm{au}}\right)^{-1}.
    \label{eq:ionize_numerical}
\end{align}
Here $m_\mathrm{b}$ is the total mass of the binary and $a$ its semimajor axis.
The passing stars have mass $m_\mathrm{p}$, number density $n$ and velocity dispersion $\sigma$, and
$\ln\Lambda$ is the Coulomb logarithm, with $\Lambda \sim \sigma^2 a/(Gm_\mathrm{p})$.
Importantly, this timescale depends on the total mass of the binary but \textit{not} on its mass ratio $q$.
   %Galactic tides, on the other hand, modify the eccentricity distribution on a characteristic timescale $t_{\mathrm{Gal}}\sim 4\ \mathrm{Gyr}\ (a/10^4\ \mathrm{au})^{-3/2}$.
%If the birth eccentricity distribution depends on mass ratio, 
%subsequent dynamical evolution may gradually weaken 
%this dependence as the eccentricity distributions, 
%if not initially thermal, evolve towards thermal. 
%Such a dependence may therefore be easier to detect in 
%younger binary populations, 
%which are more likely to retain information about their birth eccentricities. 

Heuristically, then, we expect dynamical evolution to be imprinted in the 
eccentricity distribution's dependence on binary 
{age}, while any remaining dependence on {mass ratio} 
should be telling us something about their formation processes. 
This motivates the central question of our study: 
how much of the present-day eccentricity distribution (i.e., the measured $\alpha$ values) 
reflects 
the conditions of binary formation, and how much reflects subsequent dynamical processing? 
%Examining the present-day eccentricity distribution as a function of both stellar age and 
%mass ratio can therefore help constrain both the birth eccentricity distribution 
%and its subsequent dynamical evolution. 
Observationally, stellar ages and masses are both difficult to determine directly for the 
wide binary stars, so in this study we
 use the Galactic vertical action $J_z$ as a dynamical proxy for stellar age 
 and the \textit{Gaia} $G$-band magnitude contrast $\Delta G$ 
 between binary components as a proxy for mass ratio.
 To be clear, we do not assume a 1-to-1 mapping $J_z \leftrightarrow \tau$ or 
 $\Delta G \leftrightarrow q$. Rather, we report our observations in the $(\alpha, J_z, \Delta G)$ 
 space without assuming any such connection. Then when we come to interpret 
 these results theoretically, 
 we perform forward modelling of binary dynamical evolution that 
 makes predictions in this space to be compared with the data.

The rest of this paper is organised as follows.
In Section~\ref{sec:data_methods} we describe how we infer the eccentricity distribution of a
population of binaries and how we selected our sample.
In Section~\ref{sec:results} we present our measurements of $\alpha$ as a function of $J_z$ and
$\Delta G$.
In Section~\ref{sec:discussion} we interpret these measurements with a simple model of the
dynamical evolution of wide binaries in the Galactic disc.
We summarise our results in Section~\ref{sec:summary}.

\section{Data and methods}
\label{sec:data_methods}

\subsection{Inferring the eccentricity distribution}
\label{sec:inference}
Since wide binaries have long orbital periods, we cannot directly measure their orbital eccentricities.
 \cite{Tokovinin1998} proposed that the distribution of the angle between the separation and relative 
 velocity vectors of binary components contains information about their orbital eccentricities.
  The basic idea is as follows. 
  For a circular orbit, the relative velocity is always perpendicular to the separation vector, 
  and hence the angle $\gamma_{\mathrm{3D}}=90^\circ$ throughout the orbit. 
  For an eccentric orbit, however, $\gamma_{\mathrm{3D}}$ varies with orbital phase. 
%  While $\gamma_{\mathrm{3D}}=90^\circ$ at both pericentre and apocentre, 
%  it reaches a maximum as it travels from the apocentre to the pericentre and 
%  a minimum as it travels from the pericentre to the apocentre. 
  The more eccentric a binary orbit is, the wider the range of $\gamma_{\mathrm{3D}}$ 
  values it spans.
  %, with the extrema moving further away from $90^\circ$ on either side. 
  %Near the extrema of $\gamma_{\mathrm{3D}}$, 
  %a relatively large range of orbital phases maps onto a narrow range of angles, 
  %enhancing the probability density near these values. 
  Furthermore, 
  binaries spend more time near apocentre than near pericentre,  
  making some orbital phases (hence some $\gamma_{\mathrm{3D}}$ angles) 
  more likely to be observed than others. 
  Thus, if we sample a large number of binaries
   we expect the histogram of their $\gamma_{\mathrm{3D}}$ values to have certain 
   properties depending on the eccentricity distribution from which they are drawn 
  (see figure 3a of \citealt{hwang2022}). 
Calculating the true $\gamma_{\mathrm{3D}}$ angle is difficult because the uncertainties
in distances to the individual binary components are typically larger than the line-of-sight
separation between binary components, and radial velocities are unavailable for many stars in \textit{Gaia}. Therefore, studies use the `$v$--$r$ angle', i.e., the 
angle $\gamma$ between the projected separation vector and the relative
transverse velocity vector of the binary components. 
Although projection effects mean that $\gamma$ distributions are broadened compared to those of $\gamma_\mathrm{3D}$, the basic features are preserved (see figure 3b of \citealt{hwang2022}).

Our goal is to infer the eccentricity index $\alpha$ of a binary population and thereby 
characterise its population-level eccentricity distribution. We follow the Bayesian inference framework of \citet{hwang2022}. We assume that the eccentricities of a population of binaries follow a power-law distribution, $p(e\mid\alpha)=(1+\alpha)e^\alpha$.
We use this power law only to summarise the distribution of $\gamma$ in a population by a single
number (Appendix~\ref{app:alpha}).
To relate this distribution to the projected $v$--$r$ angle $\gamma$, \citet{hwang2022} 
simulated binaries on Keplerian orbits with different eccentricities, 
orbital phases, and viewing orientations to construct a grid of probability distributions 
$p(\gamma_{\mathrm{true}}\mid e)$, where $\gamma_{\mathrm{true}}$ denotes the projected 
$v$--$r$ angle in the absence of measurement uncertainties.
The posterior distribution of $\alpha$ for a population of $N$ binaries 
is then given by,
\begin{equation}
\label{eq:posterior}
\begin{aligned}
p(\alpha\mid\{\gamma_{\mathrm{obs},i}\})
&\propto p(\alpha)\prod_{i=1}^{N}\int_0^1 \int_0^{180^\circ}
p(\gamma_{\mathrm{obs},i}\mid
\gamma_{\mathrm{true},i},\sigma_{\gamma,i})\\&\quad\times\ p(\gamma_{\mathrm{true},i}\mid e_i)
p(e_i\mid\alpha)\ d\gamma_{\mathrm{true},i}\ de_i.
\end{aligned}
\end{equation}
Here, $i$ denotes an individual binary, $\sigma_{\gamma,i}$ is its measurement uncertainty, 
and $p(\alpha)$ is the prior on the eccentricity index. We model $p(\gamma_{{\rm obs},i}\mid\gamma_{{\rm true},i},\sigma_{\gamma,i})$ as a Gaussian centred on $\gamma_{{\rm true},i}$ with width $\sigma_{\gamma,i}$, reflected at $0^\circ$ and $180^\circ$
and normalized over this interval. We adopt a uniform prior on $\alpha$ over the range $[-0.99,5]$\footnote{The power law
$p(e\mid\alpha)$ can be normalised only for $\alpha>-1$, so we consider this as a hard lower limit 
on $\alpha$ throughout the paper. In principle there is no upper limit.}
and evaluate the posterior on a grid with a spacing of $\Delta \alpha= 0.01$. 
We report the maximum a posteriori (MAP) estimate of $\alpha$ along with its 68 per cent Bayesian Highest Density Interval (HDI), which corresponds to the narrowest interval which 
encloses 68 per cent of the total posterior probability.

\begin{figure}
    \centering
    \includegraphics[width=\columnwidth]{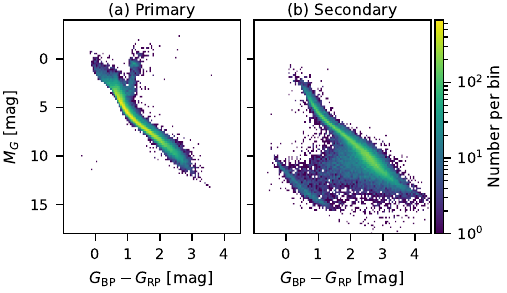}
    \caption{The colour-magnitude diagrams of the primary (left) and secondary (right) components of the wide-binary sample, consisting of 48,199 binaries with a radial velocity measurement having an uncertainty below $3\,\mathrm{km\,s^{-1}}$ for at least one component. The diagrams show 48,195 primary and 41,551 secondary components with valid colour and absolute magnitude measurements within the displayed limits.}
    \label{fig:fig1}
\end{figure}

\subsection{Sample selection}
\label{sec:sample}

We use the wide binary catalogue of \citet{hwang2022}, which builds upon the \textit{Gaia} EDR3 catalogue of
 1,817,594 candidate binaries from \citet{elbadry2021}, predominantly main-sequence binaries. 
 \citet{hwang2022} provide measurements of the projected $v$--$r$ angle ($\gamma$), along with its measurement uncertainty and the significance of the proper-motion difference between the two components.

Following \citet{hwang2022}, we apply four quality cuts to the sample.
(i) The $v$--$r$ angle $\gamma$ is measured using the difference $\Delta\mu$ between the proper
motions of the two components. Its uncertainty $\sigma_\gamma$ is set mainly by the uncertainty $\sigma_{\Delta\mu}$ in this difference.
We require $\Delta\mu/\sigma_{\Delta\mu}>3$, which ensures that $\sigma_\gamma\lesssim20^\circ$
(see equation 23 of \citealt{hwang2022}).
This cut may preferentially exclude highly eccentric binaries, because they are more likely to be
observed near apocentre, where their orbital velocities are small.
(ii) The catalogue gives for each pair the probability $\mathcal{R}$ that it is a chance alignment
of two unrelated stars rather than a bound binary.
We require $\mathcal{R}<0.1$ to reduce contamination by chance alignments.
(iii) \textit{Gaia} astrometric systematics in pairs with a small angular separation $\theta$ on the sky
can create an artificial excess near $90^\circ$ in the observed distribution of $\gamma$.
We require $\theta>1.5''$ to minimise these effects.
(iv) We require the distance $d$ to each binary to be less than 200\,pc.
This cut reduces the selection bias introduced by the requirement on proper-motion significance
and improves the completeness of the main-sequence sample.
For a detailed discussion of \textit{Gaia} systematics and the choice of quality cuts, see section 3 of
\citet{hwang2022}.
After applying these quality cuts, we retain 78,013 binaries.

In addition, to investigate whether the eccentricity distribution depends on mass ratio, 
we use the \textit{Gaia} $G$-band magnitude contrast ($\Delta G$) between binary components as a proxy, defined as $\Delta G=G_{\mathrm{secondary}}-G_{\mathrm{primary}}$, where the primary is the 
brighter component. For main-sequence binaries, since both components are approximately at 
the same distance, their flux ratio is equivalent to their luminosity ratio,
which can be related to the mass ratio through the mass-luminosity relation. The relation is not one-to-one, and it does not hold for the binaries with a white-dwarf secondary (see below).
The relation between $\Delta G$ and $q$ is much tighter than that between $J_z$ and age,
however. To estimate the mass ratios, we exclude binaries with a
white-dwarf secondary, identified by $M_G>3.25\,(G_{\rm BP}-G_{\rm RP})+9.625$ for the fainter component in Figure~\ref{fig:fig1}, and convert the absolute $G$ magnitudes of both components to masses using the main-sequence relation of \citet{Pecaut2013}. For binaries at $10^3$--$10^{3.5}$ au, the median mass ratio is 0.71 for $0.25\leq\Delta G<3$ mag and 0.16 for
$\Delta G\geq7$ mag. At a given $\Delta G$, the central 68 per cent of inferred mass
ratios span less than 0.2 dex.

We cross-match the wide binary catalogue with \textit{Gaia} DR3 \citep{GaiaCollaboration2016,GaiaCollaboration2023} to obtain radial velocities and photometry for both components. Vertical actions require six-dimensional phase-space information, so we keep the binaries in which at least one component has a radial velocity with an uncertainty below 3\,km\,s$^{-1}$. When both components have one, we use the component with the smaller uncertainty. The median uncertainty of the component that we use is 0.55\,km\,s$^{-1}$. We approximate the vertical action of the barycentre by that of this component. This neglects the orbital motion of the component about the barycentre, which is about 0.1\,km\,s$^{-1}$ for the binaries in the separation range $3\leq\log_{10}(s/\mathrm{au})<3.5$ along any one axis: 8 per cent of the median vertical speed in our lowest $J_z$ bin and less than 4 per cent in the others. These requirements leave us with a sample of 48,199 binaries. Figure~\ref{fig:fig1} shows the colour-magnitude diagrams of the primary and secondary components. Our sample is predominantly main-sequence binaries, although, as shown in Figure~\ref{fig:fig1}, a small fraction have a white-dwarf secondary.

Using the \textit{Gaia} astrometric measurements and radial velocities of the selected binary components, 
we compute their Galactocentric positions and velocities using \texttt{Astropy} \citep{Astropy2018}. We adopt a distance of the Sun from the Galactic centre $R_0=8.122$ kpc \citep{gravity2018}, a distance of the Sun from the Galactic midplane, $z_\odot=20.8$ pc \citep{Bennett2019}, and a solar Galactocentric velocity of $(12.9,245.6,7.78)$\,km\,s$^{-1}$\citep{Drimmel2018}. We then calculate the Galactic actions ($J_R$, $J_\phi$, $J_z$) using the \texttt{actionAngleStaeckel} method \citep{Binney2012,bovy2013} implemented in \texttt{galpy} \citep{Bovy2015}, adopting the \texttt{Cautun20} Galactic potential \citep{Cautun2020} and a Stäckel focal parameter of $\Delta=0.7$ in \texttt{galpy} units. We exclude 23 binaries on retrograde orbits ($J_\phi<0$), leaving a sample of 48,176 binaries.

%Motivated by the transition from subthermal to superthermal eccentricity distributions at 
%separations around $10^3$\,au \citep{hwang2022}, we focus our main analysis on binaries 
%with $3\leq\log_{10}(s/\mathrm{au})<3.5$. 

We focus most of our study on the separation bin $3\leq\log_{10}(s/\mathrm{au})<3.5$, which contains 12,234 binaries. In this bin, 
the timescale for dynamical evolution (equation~\eqref{eq:ionize_numerical}) is short enough that some dynamical evolution is expected within 10\,Gyr, but long enough that binaries born in this bin tend to stay in it for life. At much smaller separations, the timescale for dynamical evolution is much longer than a Hubble time (and these binaries truly seem to form a different population, see \citealt{hwang2022}).
At much larger separations the number of binaries is smaller, because the number of main-sequence pairs per
unit separation falls roughly as $s^{-1.6}$ beyond $s\simeq500$\,au \citep{ElBadryRix2018}.
The statistics there are therefore less reliable, and such binaries tend to increase their $s$ significantly
(or even get disrupted) within a Hubble time.
This does not mean that wider separation bins cannot be studied --- only that making 
firm statements about them is more difficult with the current data.
We therefore concentrate on 
$3\leq\log_{10}(s/\mathrm{au})<3.5$ for the most part, 
mentioning other separation bins' results only briefly.

%In addition, stellar encounters are expected to dominate over Galactic tides in this regime, 
%while the evolution of binary semimajor axes is expected to be relatively slow. 
%Since binaries formed at different separations may have different initial eccentricity 
%distributions, restricting our analysis to this narrow separation range 
%allows us to investigate how the eccentricity distribution varies with $J_z$ 
%while controlling for present-day projected separation and potentially limiting
% differences in birth separations.

\section{Results}
\label{sec:results}
\begin{figure*}
    \centering
    \includegraphics[width=0.9\linewidth]{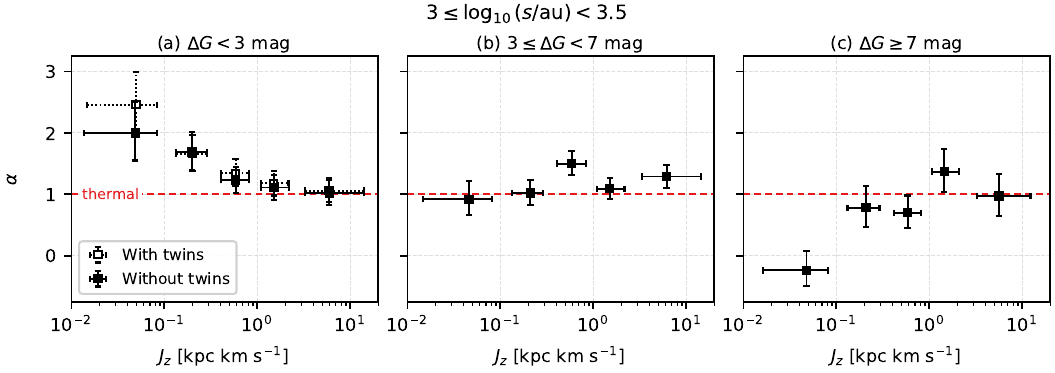}
    \caption{Eccentricity index $\alpha$ against Galactic vertical action $J_z$ for binaries with projected
    separations of $10^3$--$10^{3.5}$\,au, in three bins of $\Delta G$. Filled squares show the binaries without twins and, in panel (a), open squares with dotted bars the binaries
    including twins ($\Delta G<0.25$\,mag).
    Vertical bars show the 68 per cent highest-density intervals of $\alpha$, and horizontal bars span the
    central 68 per cent of the $J_z$ values in each bin.
    Red dashed lines mark the thermal distribution, $\alpha=1$.
    The bars and the red dashed lines have the same meanings in all later figures of $\alpha$.}
    \label{fig:fig2}
\end{figure*}
\begin{figure*}
    \centering
    \includegraphics[width=0.9\linewidth]{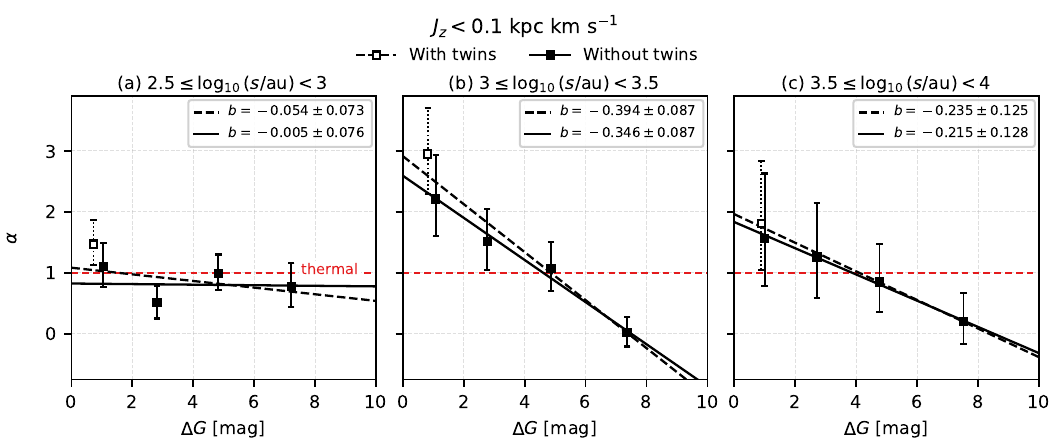}
    \caption{Eccentricity index $\alpha$ against $\Delta G$ for binaries with $J_z<0.1$\,kpc\,km\,s$^{-1}$, in
    three ranges of projected separation and the four $\Delta G$ bins of Table~\ref{tab:table2}.
    Squares are as in Figure~\ref{fig:fig2}.
    Solid and dashed lines show the straight lines $\alpha=a+b\Delta G$ that we fitted by weighted least squares
    to the values without and with twins, and the legends give their slopes $b$ in mag$^{-1}$.}
    \label{fig:fig3}
\end{figure*}

\begin{table*}
\centering
\small

\setlength{\tabcolsep}{5pt}
\renewcommand{\arraystretch}{1.5}

\caption{
Inferred eccentricity indices for binaries with
$3\leq\log_{10}(s/\mathrm{au})<3.5$.
The values correspond to the maximum posterior estimates
of $\alpha$, with uncertainties given by the
68 per cent highest-density intervals.
$N$ denotes the number of binaries in each subsample.
$J_z$ is expressed in $\mathrm{kpc\,km\,s^{-1}}$
and $\Delta G$ in magnitudes.
}
\label{tab:alpha_main}
\begin{tabular}{lcccccccc}
\hline
\hline

\multicolumn{9}{c}{
$3\leq\log_{10}(s/\mathrm{au})<3.5$
}\\[4pt]

\hline

$J_z$ &
\multicolumn{2}{c}{\shortstack{$0\leq\Delta G<3$\\(with twins)}} &
\multicolumn{2}{c}{\shortstack{$0.25\leq\Delta G<3$\\(without twins)}} &
\multicolumn{2}{c}{$3\leq\Delta G<7$} &
\multicolumn{2}{c}{$\Delta G\geq7$} \\

\cline{2-9}

& $N$ & $\alpha$
& $N$ & $\alpha$
& $N$ & $\alpha$
& $N$ & $\alpha$ \\

\hline
0--0.1 & 406 & $2.46^{+0.54}_{-0.48}$ & 352 & $2.00^{+0.51}_{-0.45}$ & 460 & $0.92^{+0.30}_{-0.26}$ & 150 & $-0.24^{+0.31}_{-0.25}$ \\
[6pt]0.1--0.336 & 705 & $1.66^{+0.30}_{-0.28}$ & 643 & $1.69^{+0.32}_{-0.30}$ & 935 & $1.02^{+0.22}_{-0.20}$ & 315 & $0.78^{+0.36}_{-0.32}$ \\
[6pt]0.336--0.950 & 1,111 & $1.34^{+0.23}_{-0.20}$ & 993 & $1.23^{+0.22}_{-0.21}$ & 1,525 & $1.50^{+0.21}_{-0.18}$ & 485 & $0.70^{+0.28}_{-0.25}$ \\
[6pt]0.950--2.668& 1,111 & $1.18^{+0.20}_{-0.20}$ & 999 & $1.11^{+0.21}_{-0.20}$ & 1,500 & $1.09^{+0.18}_{-0.16}$ & 464 & $1.37^{+0.37}_{-0.33}$ \\
[6pt]$>2.668$ & 1,111 & $1.06^{+0.20}_{-0.18}$ & 1,019 & $1.02^{+0.21}_{-0.19}$ & 1,597 & $1.29^{+0.18}_{-0.18}$ & 359 & $0.97^{+0.36}_{-0.32}$ \\

\hline
\end{tabular}

\end{table*}
For binaries with projected separations in the range $10^3$--$10^{3.5}$\,au, we divide the sample into three $\Delta G$ bins: 0--3 mag, 3--7 mag, and $>7$ mag. We then divide the $J_z$ distribution of binaries in the $\Delta G=0$--3 mag bin into four equal-count bins, with the lowest $J_z$ bin spanning 0--0.336\,kpc\,km\,s$^{-1}$. We further subdivide this lowest bin at $J_z=0.1$\,kpc\,km\,s$^{-1}$ to isolate the dynamically coldest population, which may retain more information about the birth eccentricity distribution. We use the same $J_z$ bin boundaries across different $\Delta G$ bins to ensure a fair comparison and retain these boundaries throughout our analysis unless otherwise stated. We then infer the population eccentricity index $\alpha$ for the binaries in each $J_z$ bin across different $\Delta G$ bins. We also repeat the analysis after excluding binaries with $\Delta G<0.25$ mag, which corresponds to the twin population ($q\gtrsim0.95$) for main-sequence binaries \citep{ElBadry2019twins,hwang2022b}.

The main results are summarised in Figure~\ref{fig:fig2} and Table~\ref{tab:alpha_main}.
We now discuss each panel of Figure~\ref{fig:fig2} in turn. We do not perform any theoretical interpretation here, deferring this to the Discussion.

(a) $\Delta G<3$ mag. Binaries in this $\Delta G$ bin have components of similar brightness 
and hence comparatively precise astrometry relative to binaries in the other 
$\Delta G$ bins. Consequently, their $\gamma$ measurements are also precise with a median uncertainty of $2.36^\circ$, compared with $4.75^\circ$ and $7.71^\circ$ for $3\leq\Delta G<7$ and $\Delta G\geq7$, respectively. The figure shows the $\alpha$ versus $J_z$ trend with and without twins. We find that $\alpha$ declines strongly with increasing $J_z$ in this regime. The lowest $J_z$ bin is strongly superthermal, whereas the highest $J_z$ bin is close to thermal. The lowest-$J_z$ bin contains a relatively larger fraction of twin binaries compared to the other $J_z$ bins. Removing the twins reduces the point estimate of $\alpha$ from 2.46 to 2.00, but does not eliminate the declining trend with $J_z$.
This is consistent with \citet{hwang2022b}, who found that wide twins have much more eccentric orbits than other wide binaries, which suggests that they form differently.
We thus recover the known result for twins and extend it: not only the twins but also the other binaries with $\Delta G<3$\,mag are superthermal at low $J_z$.

%The observed decline in $\alpha$ with $J_z$ is qualitatively consistent with the eccentricity evolution
% towards a thermal distribution predicted by \citet{Hamilton2024}. However, the thermalization 
%timescales due to stellar encounters are typically long in this separation 
%range \citep[][equation 16]{Hamilton2024}. Hence, the decline from $\alpha=2.11^{+0.55}_{-0.47}$ 
%in the lowest-$J_z$ bin to $\alpha=1.09^{+0.21}_{-0.21}$ in the highest-$J_z$ bin cannot be
%explained by dynamics alone, assuming that all binaries share the same birth eccentricity distribution.

(b) $3\leq\Delta G<7$ mag. Binaries in this $\Delta G$ range show an approximately flat $\alpha(J_z)$ trend,  
with most MAP estimates
close to being thermal. In particular, the lowest-$J_z$ bin is consistent with a thermal 
distribution, in contrast to the corresponding $\Delta G<3$ mag bin 
which is strongly superthermal. 
However, the third $J_z$ bin shows a superthermal distribution ($\alpha=1.50^{+0.21}_{-0.18}$), while the highest-$J_z$ bin is also mildly superthermal ($\alpha=1.29^{+0.18}_{-0.18}$).

(c) $\Delta G\geq7$ mag. In contrast to the $\Delta G<3$ mag sample,
the lowest-$J_z$ bin in this regime is strongly subthermal, with $\alpha=-0.24^{+0.31}_{-0.25}$.
% This value changes by less than its uncertainty when we remove the white-dwarf secondaries, raise the threshold of cut (i) or require more precise radial velocities (see below). %(\S\ref{sec:sample}).
%This is particularly interesting because these binaries are expected to lie at the opposite extreme
% of the mass-ratio distribution from the twins, suggesting a `birth' eccentricity distribution 
% that depends on the mass ratio. 
At higher $J_z$, $\alpha$ approaches the thermal value.

The strong dependence of the inferred $\alpha$ on $\Delta G$ in the lowest $J_z$ bin is also apparent in the observed $\gamma$ distributions (Appendix\cmark{~}\ref{sec:obs_gamma_dist}, Figure\cmark{~}\ref{fig:C_1}). In the lowest $J_z$ bin, the observed $\gamma$ distribution decreases towards $90^\circ$ for $0.25\leq\Delta G<3$ mag, is approximately flat for $3\leq\Delta G<7$ mag, and rises towards $90^\circ$ for $\Delta G\geq7$ mag. With increasing $J_z$, the $\gamma$ distributions in the two outer $\Delta G$ bins become progressively flatter, approaching the approximately uniform $\gamma$ distribution expected for a thermal eccentricity distribution.

\begin{table}
\centering
\small
\setlength{\tabcolsep}{5pt}
\caption{Numbers of binaries with $J_z<0.1\,\mathrm{kpc\,km\,s^{-1}}$ in the four $\Delta G$ bins of
Figure~\ref{fig:fig3}, in three ranges of projected separation.
Numbers in brackets exclude the twins ($\Delta G<0.25$\,mag), all of which lie in the first bin.}
\label{tab:table2}
\begin{tabular}{lcccc}
\hline
& \multicolumn{4}{c}{$\Delta G$ [mag]} \\
$\log_{10}(s/\mathrm{au})$ & $0$--$1.903$ & $1.903$--$3.804$ & $3.804$--$5.963$ & $\geq5.963$ \\
\hline
$2.5$--$3$ & 395 (304) & 385 & 440 & 260 \\
$3$--$3.5$ & 254 (200) & 254 & 254 & 254 \\
$3.5$--$4$ & 114 (92) & 115 & 119 & 112 \\
\hline
\end{tabular}

\end{table}
In Figure~\ref{fig:fig2} the eccentricity distribution of the dynamically coldest binaries
($J_z<0.1$ \,kpc\,km\,s$^{-1}$) depends strongly on $\Delta G$. 
%Since $J_z$ serves as a proxy for age, binaries in this population are expected to be younger 
%on average than those at higher $J_z$. 
%Most of these binaries are probably older than 1\,Gyr (Section~\ref{sec:model}). 
%They may still retain much of their birth eccentricity distribution, because at these separations 
%stellar encounters and the Galactic tide change the eccentricity distribution only over several Gyr.
To characterise this dependence we selected only those 1,016 binaries 
with $3\leq\log_{10}(s/\mathrm{au})<3.5$ and $J_z<0.1$ \,kpc\,km\,s$^{-1}$, of which 54 are twins. 
We divide this sample into four equal count bins in $\Delta G$ and infer $\alpha$ for each 
bin. We then apply the same bin boundaries to binaries in the separation 
ranges $2.5\leq\log_{10}(s/\mathrm{au})<3$ and $3.5\leq\log_{10}(s/\mathrm{au})<4$ 
and repeat the analysis. 
The results are shown in Figure~\ref{fig:fig3}, and Table~\ref{tab:table2} gives the numbers
of binaries in each bin.\footnote{In Appendix~\ref{app:dG} we also fit $\alpha=a+b\Delta G$ to the
individual binaries, without binning them in $\Delta G$, and find the same trends.}
As shown in Figure~\ref{fig:fig3}b, for binaries with $3\leq\log_{10}(s/\mathrm{au})<3.5$, 
the inferred eccentricity index shows a clear decreasing trend with increasing $\Delta G$, roughly
 $\alpha=a+b\Delta G$ with $(a, b) = (2.9,-0.39\,{\rm mag}^{-1})$
when we include the twins and $(a, b) = (2.6,-0.35\,{\rm mag}^{-1})$
when we do not.
We fitted these straight lines to the four values of $\alpha$ by weighted least squares, with half the
width of each 68 per cent HDI as the uncertainty in $\alpha$.
A similar trend seems to persist in the $3.5\leq \log_{10}(s/\mathrm{au})<4$ bin as well
(Figure~\ref{fig:fig3}c), although the dependence on $\Delta G$
is not quite as steep. By contrast, 
binaries with $2.5\leq\log_{10}(s/\mathrm{au})<3$ (Figure~\ref{fig:fig3}a) have a (roughly) thermal 
distribution, $\alpha\simeq1$, regardless of $\Delta G$, consistent with \cite{hwang2022}, who reported a transition from thermal to superthermal eccentricity distribution at separations $\sim 10^3$\,au.

We also tested the sensitivity of the slopes to different $\Delta G$ boundaries and found that $2.5\leq\log_{10}(s/\mathrm{au})<3$ and $3\leq\log_{10}(s/\mathrm{au})<3.5$ have relatively stable slope
estimates. For $3.5\leq\log_{10}(s/\mathrm{au})<4$, the degree of overlap between the 68 per cent HDIs of $\alpha$ in the first
and last $\Delta G$ bins seems to be sensitive to the binning choices, although the fitted slopes remain consistent within their uncertainties. For example, fixed $\Delta G$ bins of 0--3, 3--5, 5--7 and $\geq7$\,mag give slopes of $0.012\pm0.082$
($-0.042\pm0.078$), $-0.345\pm0.079$ ($-0.389\pm0.080$) and $-0.323\pm0.135$\,mag$^{-1}$ ($-0.328\pm0.128$)
for the separations $2.5\leq\log_{10}(s/\mathrm{au})<3$, $3\leq\log_{10}(s/\mathrm{au})<3.5$ and $3.5\leq\log_{10}(s/\mathrm{au})<4$ respectively, for the sample without twins (with twins).

We recover these trends in several additional tests. When we order the binaries by ages inferred from their stellar parameters instead of by $J_z$ (Appendix~\ref{app:ages}, Figure~\ref{fig:A_3}), $\alpha$ evolves from superthermal towards thermal with increasing inferred age for $0.25\leq\Delta G<3$ mag, and from subthermal towards thermal for $\Delta G\geq7$ mag. The strong dependence of $\alpha$ on $\Delta G$ in the kinematically coldest population is also recovered when we use transverse velocity $V_T$ instead of $J_z$, without imposing a radial-velocity requirement (Appendix~\ref{app:vt}, Figure~\ref{fig:B_1}). To test whether the trends depend on how far the binaries oscillate radially in the Galaxy, we also split each $\Delta G$ bin at the median radial action $J_R$ of the binaries with
$0.25\leq\Delta G<3$\,mag, and plot $\alpha$ against $J_z$ for each half in Figure~\ref{fig:C2_4} (Appendix~\ref{app:jz}).
In both halves $\alpha$ falls with $\Delta G$ in the lowest $J_z$ bin, and for $0.25\leq\Delta G<3$\,mag the fall of $\alpha$ with $J_z$ is steeper in the half with low $J_R$. The trends therefore persist in each of these tests, whichever proxy for age we use.

We also tested whether the main trends are sensitive to our sample selection criteria. Our sample is not complete. The requirement of a radial velocity with an uncertainty below 3\,km\,s$^{-1}$ removes 38 per cent of the binaries that pass the quality cuts.
Fortunately, we do not need a complete sample for our measurements or for the models of
Section~\ref{sec:discussion}, as long as the selection does not depend on eccentricity or on age.
We have several reasons to believe that it does not.
For instance, \citet{hwang2022} passed simulated binaries through the quality cuts and found that these cuts raise a
true value of $\alpha\simeq1.2$ to about 1.3 and leave a thermal distribution thermal (their section 3.6).
The cuts may therefore slightly exaggerate departures from a thermal distribution.
Also, raising the threshold of cut (i) from 3 to 5 changes $\alpha$ by less than its uncertainty in every bin of
Table~\ref{tab:alpha_main}.
The same is true when we remove the binaries with a white-dwarf secondary, identified using the colour--magnitude criterion mentioned in \S\ref{sec:sample}.
These binaries need particular care: the mass loss that made the white dwarf also changes the orbit, and at a given
separation they have lower eccentricities than main-sequence pairs \citep{hwang2025white}.
Such binaries comprise 5 per cent of the population at $3\leq\log_{10}(s/\mathrm{au})<3.5$ and 17 per cent of those that also have $\Delta G\geq7$\,mag, but only
9 per cent of those in the lowest $J_z$ bin, where $\alpha$ is lowest.
Removing them lowers $\alpha$ in that bin from $-0.24$ to $-0.33$, and so does removing every secondary within 3\,mag of this line.
When we keep only the components with a radial-velocity uncertainty below 1\,km\,s$^{-1}$, $\alpha$ in the lowest $J_z$ bin falls from 2.00 to 1.61 for $0.25\leq\Delta G<3$\,mag and from $-0.24$ to $-0.31$ for $\Delta G\geq7$\,mag, both within the uncertainties, so the dependence of $\alpha$ on $\Delta G$ at low $J_z$ remains.
In addition, within each $\Delta G$ bin the distributions of magnitude and colour change little with $J_z$ (Appendix~\ref{app:jz}).
In summary, we do not believe that selection effects are significantly biasing our measurements.
The one exception that we found is at low transverse velocity, where $\alpha$ depends on whether we require a radial velocity (Appendix~\ref{app:vt}); we return to this in \S\ref{sec:origins}.

\section{Discussion}
\label{sec:discussion}

In this section we interpret our measurements with simple dynamical
models of wide binary evolution.
In \S\ref{sec:model} we describe our fiducial model in which we believe
the physics is well understood. We show in \S\ref{sec:fiducial}
that it under-predicts the rate at which binaries evolve
in the lowest $J_z$ bins compared to the measurements of \S\ref{sec:results}.
We then examine two changes to our model that bring it closer to our measurements.
In \S\ref{sec:faster} we boost the encounter rate above that of the Galactic field for the first few Gyr of each
binary's life, and in \S\ref{sec:timedep} we let the birth distribution change with time.
In \S\ref{sec:seps} we present measurements in other separation bins, and in
\S\ref{sec:origins} we discuss the questions that our study leaves open.

\subsection{Dynamical evolution of wide binaries}
\label{sec:model}

\begin{figure}
\centering
\includegraphics[width=0.95\columnwidth]{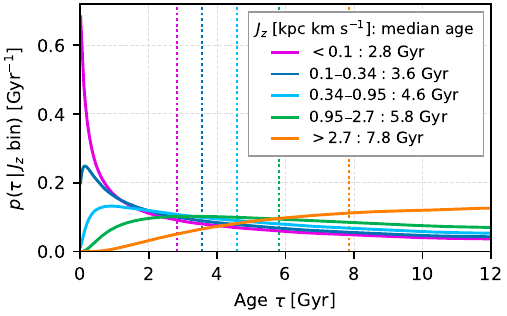}
\caption{Ages of the binaries at $10^3$--$10^{3.5}$\,au in each $J_z$ bin of
Section~\ref{sec:results}, predicted by the model of \S\ref{sec:model}.
Each curve is the distribution of ages $p(\tau\,|\,j)$ of the binaries in one $J_z$ bin $j$
(Appendix~\ref{app:predict}), normalised so that the area under it is 1.
The legend gives the colour of each bin.
Dotted lines mark the median ages, which the legend also gives.}
\label{fig:jz_age}
\end{figure}

To interpret Figure~\ref{fig:fig2} we need to know how 
wide binaries evolve dynamically as they age, 
and to relate their ages to $J_z$.
To do this, we use a dynamical model which we describe in full in
Appendix~\ref{app:model}. There we also test each choice that we made in the model, among them the distribution of semimajor axes at
birth, the birth rate, the treatment of encounters as a diffusion process, the vertical excursions and radial epicycles
of the binaries, and the selection of model binaries by semimajor axis.
For our fiducial model in which the passing star encounters are sampled at the present day 
Galactic field rates, none of these tests changes the predicted values of $\alpha$ by
more than 0.1.

Briefly, our model follows populations of binaries born over the last 12\,Gyr, each with an
eccentricity drawn from the power law $p(e)\propto e^{\alpha_{\rm i}}$.
We call $\alpha_{\rm i}$ the \emph{birth index}, to distinguish it from the index $\alpha$ that
we measure today.
We then evolve each binary by Monte Carlo sampling of the kinetic prescription developed
by \citet{Modak2023} and \citet{Hamilton2024}.
Firstly, passing stars give the binary random weak kicks to its relative 
velocity vector, changing both its semimajor axis
and its eccentricity.
Secondly, the tidal field of the Galactic disc (computed in an orbit-averaged approximation) 
changes the eccentricity of the binary smoothly,
without changing its semimajor axis.
The first effect is expected to be dominant in our separation range of interest
\citep{Hamilton2024}, and operates on the timescale of equation~\eqref{eq:ionize}.
The encounter rate is proportional to the local number density of stars, and the tidal torque to the
local mass density of the disc.
In our model both densities depend only on the guiding radius $R_\mathrm{g}$ of the binary's
barycentre, and in the same way.
Stars migrate radially through the disc over their lifetimes and spend time in regions denser
than the solar neighbourhood, so we let $R_\mathrm{g}$ random walk according to the measurement
of \citet{Modak2026}.
We do not follow the vertical motion of the binaries' barycentres explicitly.
Instead, at the end of our models we give each model binary a vertical action drawn from the
distribution expected for stars of its age.
For this we use the vertical velocity dispersions of \citet{Sharma2021} and the potential of
\citet{Cautun2020}, with which we compute $J_z$ for our sample.
At a given age we therefore treat the vertical heating of a binary's barycentre and the evolution of
its orbit as independent.\footnote{The barycentre is heated mainly by fluctuations of the Galactic potential
on scales much larger than a wide binary, such as those due to giant molecular clouds, spiral arms and passing dwarf
galaxies.
These fluctuations accelerate both stars of a binary almost equally.
The orbit of a wide binary is changed mainly by stars that pass within a few separations of it, and the combined
effect of these encounters on the barycentre is negligible compared with that of the large-scale fluctuations.}
%Binaries with large $J_z$ do spend more time far from the midplane, where encounters are rarer.
%We left this effect out of our model, and in Appendix~\ref{app:clocks} we show that including it
%changes the curves of $\alpha$ against $J_z$ by at most 0.05.

\begin{figure*}
\includegraphics[width=0.9\textwidth]{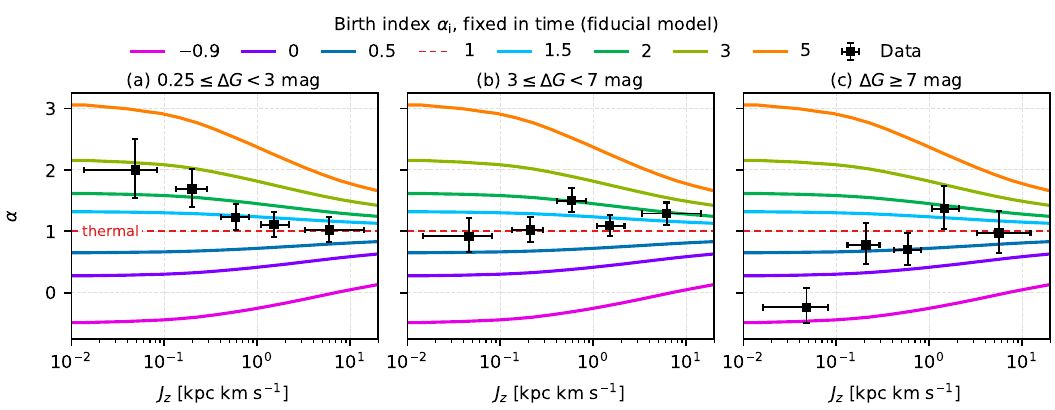}
\caption{As Figure~\ref{fig:fig2} without the twins (Table~\ref{tab:alpha_main}), with our fiducial model of
eccentricity evolution (\S\ref{sec:fiducial}), in which the encounter rate is that of the Galactic field today.
Coloured lines are for populations whose birth index $\alpha_{\rm i}$ does not change with time.
The legend gives $\alpha_{\rm i}$ for each line, and the lines are the same in every panel.
The line for $\alpha_{\rm i}=1$ lies on the red dashed lines.}
\label{fig:fan}
\end{figure*}

One input to our model is an assumed rate at which binaries were born over the last 12\, Gyr.
\citet{AumerBinney2009} found that the star formation rate per unit area of the disc near the Sun has fallen by a factor of 2 to 7 over the life of the disc.
Old stars are vertically hotter than young stars, so a smaller share of them lies within 200\,pc of
the Sun, the volume of our sample.
We take a birth rate per unit area inversely proportional to this share, so that our volume would hold
equal numbers of binaries of each age if none had been disrupted.
This birth rate was 3.6 times higher 12\,Gyr ago than it is today, inside the measured range.
The resulting age distribution for stars in each $J_z$ bin is summarised in
Figure~\ref{fig:jz_age}.
Stars of the same age have a broad range of $J_z$, so each $J_z$
bin contains stars of almost every age: $J_z$ orders the binaries by age only on average.
It can reveal a dependence on age, but it blurs any sharp feature in age.
In our model only 28 per cent of the binaries with $J_z<0.1\Jzu$ are younger than 1\,Gyr.
For this reason the precise birth rate matters little.
In our fiducial model (\S\ref{sec:fiducial}), in which the encounter rate with passing stars is the one expected in
the Galactic field today, every other birth rate that we tried changes the curves of $\alpha$ against $J_z$ by less
than 0.05.
In this model, the birth rate that best reproduces the number of our binaries in each $J_z$ bin is also close to ours
(Appendix~\ref{app:birth}).
Models with drastically different encounter rates (\S\ref{sec:faster}) are more sensitive to the birth rate, but
their $\alpha$ versus $J_z$ curves still change by less than the error bars of our measurements.
We then select the model binaries with semimajor axes between $10^3$ and $10^{3.5}$\,au in
each $J_z$ bin of Table~\ref{tab:alpha_main}, compute their projected $v$--$r$ angles, and
measure $\alpha$ from these angles with the method of Section~\ref{sec:inference}.
In other words, we predict with our model the same quantity that we measure, binned in the
same way.

\subsubsection{The fiducial model}
\label{sec:fiducial}

In our fiducial model we evolve the binaries according
to the present day number density and velocity dispersion of stars in the Milky Way.
The only free parameter in the model is then the birth index $\alpha_\mathrm{i}$.
In Figure~\ref{fig:fan} we compare the predictions of the fiducial model for various $\alpha_\mathrm{i}$
 with the data points from
Figure~\ref{fig:fig2} (excluding twins).
Qualitatively there is agreement, in the sense that
$\alpha \to 1$ as $J_z\to \infty$ (since eccentricities always thermalize eventually).
However, there is clear quantitative disagreement:
in both panel (a) and panel (c), the predicted $\alpha$ versus $J_z$ curves are much shallower
than the observed ones. That is, the model evolves much more slowly than the data require.

\subsubsection{Boosted encounter rate}
\label{sec:faster}

\begin{figure*}
\includegraphics[width=0.9\textwidth]{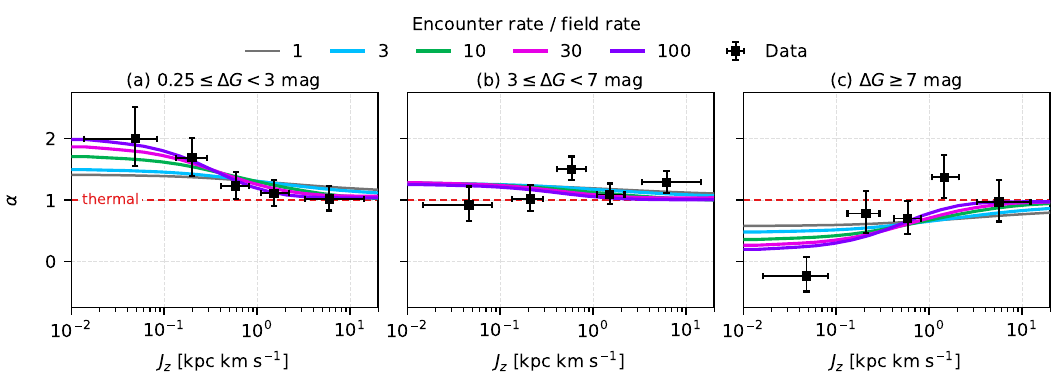}
\caption{As Figure~\ref{fig:fan}, for encounters that are more frequent than in the Galactic field at all
ages (\S\ref{sec:faster}).
Coloured lines show our model with the encounter rate multiplied by the number given in the
legend and the Galactic tide unchanged.
The grey line is for our fiducial model, with the encounter rate of the Galactic field today.
Each line is for the birth index in that $\Delta G$ bin, fixed in time, that comes closest to
the measurements.}
\label{fig:const}
\end{figure*}

The simplest interpretation of the disagreement in Figure~\ref{fig:fan} is that binaries evolve much faster than
our model predicts.
To see how much faster, we multiplied the rate of encounters (but not the Galactic tide) by a constant factor of 3,
10, 30 or 100, and chose the birth index in each $\Delta G$ bin again.
We plot the results in Figure~\ref{fig:const}.
With a factor of 10 to 30 the predicted curves pass close to our measurements in panel (a).
In panel (c), even with a factor of 100 the predicted curve stays above our lowest-$J_z$ measurement.
Encounters this frequent at all ages are implausible, because the number density and velocity dispersion of stars
near the Sun are well measured.
They would also disrupt most wide binaries, and old binaries more than young ones.
With the rate multiplied by 100, only 6 per cent of all the wide binaries ever formed would still be bound today.
Only 2 per cent of our binaries would then lie in the highest $J_z$ bin, whereas we observe 25 per cent.
To make up for such losses, the birth rate per unit area must have fallen far more steeply than the star formation
rate near the Sun, which fell by a factor of 2 to 7 over the life of the disc \citep{AumerBinney2009}.
For encounter rates 10 and 30 times that of the field the birth rate must have fallen by factors of tens to hundreds,
and for a rate 100 times that of the field no birth rate that falls exponentially with time suffices
(Appendix~\ref{app:birth}).

Encounters are more frequent than in the field while a binary is still in its birth cluster or association, whose
stars are much denser than those of the field (individual kicks are also stronger, because 
stars move more slowly relative to each other --- see equation \eqref{eq:ionize}).
However, no excess of encounters that is confined to the first few hundred Myr of each binary's life can reproduce the trend of $\alpha$
with $J_z$, whatever its form.
The reason is that an excess that ends early acts on every binary before we observe it.
Even in the lowest $J_z$ bin, most binaries in our model are older than 1\,Gyr, and only 13 per cent are younger
than 300\,Myr (Figure~\ref{fig:jz_age}).
An excess that lasts for tens or hundreds of Myr therefore changes these binaries as much as it changes the
binaries at high $J_z$.
Such an excess alters the eccentricity distribution with which binaries leave their birth environment, but not the
trend with $J_z$.
Nor can such an excess have been very strong.
If it delivers more encounters than our fiducial model does in about 30\,Gyr, more than twice the age of the
disc, even binaries born with $\alpha_{\rm i}=10$,
the most eccentric birth distribution in our model, end up below the error bar of our lowest-$J_z$ measurement for
$0.25\leq\Delta G<3$\,mag.
To explain the trend of $\alpha$ with $J_z$, the encounter rate must instead stay well above that of our fiducial
model for a few Gyr.
As an illustration, we multiplied the encounter rate of our fiducial model by
\begin{equation}
    1+\frac{B}{1+\tau/\tau_{\rm d}},
    \label{eq:boost}
\end{equation}
where $\tau$ is the age of the binary and $B$ and $\tau_{\rm d}$ are positive constants.
We took $B=60$ and $\tau_{\rm d}=300$\,Myr.
The factor is then about 15 at an age of 1\,Gyr and about 3 at an age of 10\,Gyr.
We chose the birth index in each $\Delta G$ bin again, and found that the predictions of this model come close to our
measurements (blue lines in Figures~\ref{fig:youngold} and~\ref{fig:seps}).
When we set the factor of equation~\eqref{eq:boost} to 1 for binaries older than 300\,Myr, the predicted curves
returned to those of our fiducial model.
They did so even when we raised $B$ so that the total number of extra encounters stayed the same.
When we set the factor to 1 beyond an age of 1\,Gyr instead, the curves moved only part of the way towards our
measurements.
We also tried an excess of encounters that decays exponentially with age.
With a decay time of up to about 200\,Myr we again obtained the curves of our fiducial model
(Appendix~\ref{app:faster}).
However, few birth environments last for a few Gyr.
About half of the clusters near the Sun become unbound within about 10\,Myr \citep{Lamers2005}, and a bound cluster
survives for 1 to 3\,Gyr only if its mass is about $10^4\,\mathrm{M}_\odot$ \citep{Lamers2005,Almeida2025}.

In short, we do not regard a boosted encounter rate as a satisfactory explanation of our measurements.
The boost must last longer than most birth environments survive, and we imposed it by hand rather than from a
physical model of such an environment.

\subsubsection{Time-dependent birth distribution}
\label{sec:timedep}

\begin{figure*}
\includegraphics[width=0.9\textwidth]{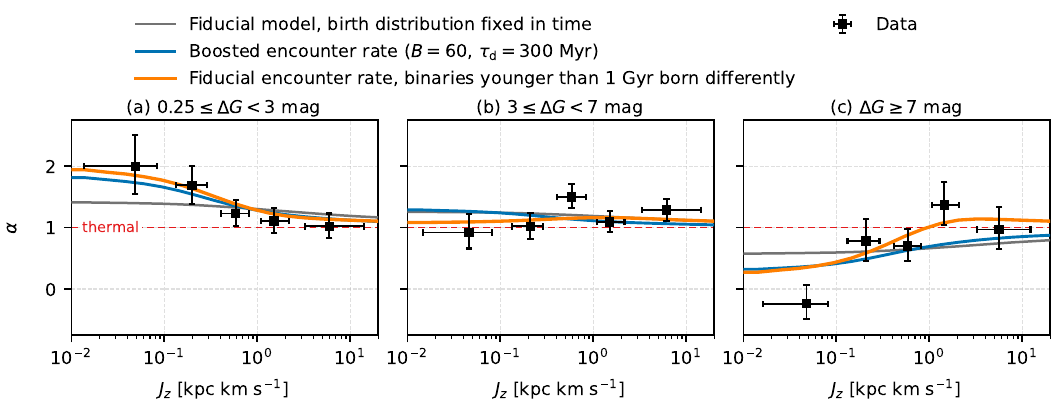}
\caption{As Figure~\ref{fig:fan}, for a birth distribution that changes with time
(\S\ref{sec:timedep}).
Orange lines show our model with the encounter rate of the Galactic field today and the birth index of
equation~\eqref{eq:youngold}, in which binaries younger than 1\,Gyr have one birth index for each
$\Delta G$ bin and older binaries share a single birth index.
For comparison, blue lines show our model with the encounter rate boosted by
equation~\eqref{eq:boost} ($B=60$ and $\tau_{\rm d}=300$\,Myr) and grey lines show our fiducial model, with
the encounter rate of the Galactic field today, each for a birth index fixed in time.}
\label{fig:youngold}
\end{figure*}

Another possibility is that all wide binaries are in fact not born with a single birth index $\alpha_\mathrm{i}$,
 but instead that birth index changes with time.
We tried several models in which the birth index changes with time.
The simplest of them that could approximately explain our measurements is as follows.
Binaries older than an age $\tau_{\rm t}$ share a single birth index whatever their $\Delta G$,
while binaries younger than $\tau_{\rm t}$ have a separate birth index in each $\Delta G$ bin:
\begin{equation}
    \alpha_{\rm i}(\tau,\Delta G)=
    \begin{cases}
        \alpha_{\rm young}(\Delta G), & \tau<\tau_{\rm t},\\
        \alpha_{\rm old}, & \tau\geq\tau_{\rm t}.
    \end{cases}
    \label{eq:youngold}
\end{equation}
Here $\alpha_{\rm young}$ takes one value in each of our three $\Delta G$ bins. 
We set the encounter
rate equal to that of our fiducial model, which is set by the Galactic field today
(i.e., we do not impose any rate boost like in \S\ref{sec:faster}).
The version of this model that comes closest to our measurements has $\tau_{\rm t}=1$\,Gyr and
$\alpha_{\rm old}=1.4$.
For $\Delta G<3$\,mag it has $\alpha_{\rm young}\simeq5$, and for $3\leq\Delta G<7$\,mag it has
$\alpha_{\rm young}=0.8$.
For $\Delta G\geq7$\,mag it has $\alpha_{\rm young}=-0.9$, the lowest birth index in our model.
Because this model has the same encounter rate and birth rate as our fiducial model, it predicts the same
number of binaries in each $J_z$ bin, close to the number that we observe (Appendix~\ref{app:birth}).
In Figure~\ref{fig:youngold} we plot the predictions of this best-fit model as orange lines, 
alongside the results of the model with the boosted encounter rate of \S\ref{sec:faster}
 with $B=60$ and $\tau_{\rm d}=300$\,Myr.
For $\Delta G<3$\,mag the two models give nearly the same curve, though 
they differ for the old binaries with fainter companions.
The curves of neither model reach our lowest-$J_z$ measurement for $\Delta G\geq7$ \,mag.
Either this measurement lies low by chance, or the lowest $J_z$ bin contains more young binaries
than our model of vertical heating assigns to it.

We do not regard the model of equation~\eqref{eq:youngold} as physically motivated, since 
we know of nothing special about star formation near the Sun 1\,Gyr ago that would have
changed how wide binaries form. Moreover, in this model old binaries of every mass ratio must share one birth distribution while young binaries
need very different ones depending on their mass ratio,
 and we know of no mechanism that would produce this.

\subsection{Other separation bins}
\label{sec:seps}

\begin{figure*}
\includegraphics[width=0.9\textwidth]{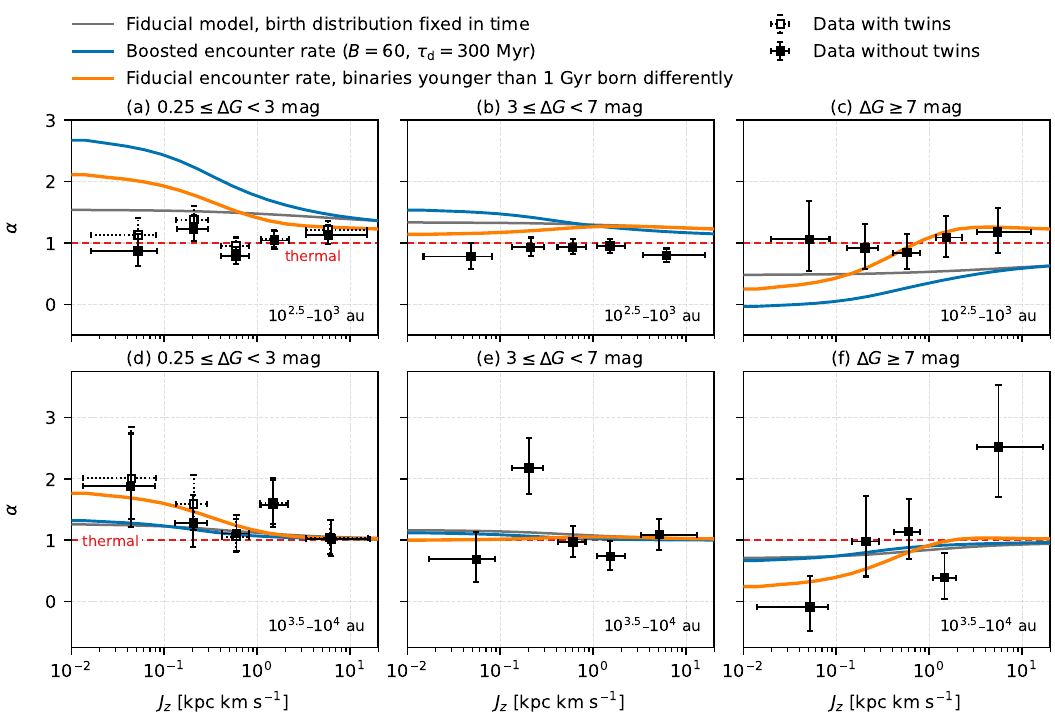}
\caption{As Figure~\ref{fig:fig2}, for binaries with
$2.5\leq\log_{10}(s/\mathrm{au})<3$ (panels a--c) and $3.5\leq\log_{10}(s/\mathrm{au})<4$ (panels d--f).
Open squares in panels (a) and (d) include the twins.
Grey lines show our fiducial model, with the encounter rate of the Galactic field today and a birth index
fixed in time.
Blue lines show our model with the encounter rate boosted by equation~\eqref{eq:boost} ($B=60$ and
$\tau_{\rm d}=300$\,Myr), and orange lines our model with the birth index of
equation~\eqref{eq:youngold}.
Each model has the birth indices that come closest to our measurements at $10^3$--$10^{3.5}$\,au
(Figures~\ref{fig:const}--\ref{fig:youngold}).}
\label{fig:seps}
\end{figure*}

So far we focused on binaries with projected separations $10^{3}$--$10^{3.5}$\,au.
However we also performed the same measurements in the bins
of $10^{2.5}$--$10^3$\,au and $10^{3.5}$--$10^4$\,au.
In Figure~\ref{fig:seps} we plot the key results.
We also plot the predictions of three of our models: our fiducial model (\S\ref{sec:fiducial}), the model
with the boosted encounter rate of \S\ref{sec:faster} with $B=60$ and $\tau_{\rm d}=300$\,Myr, and the model of
\S\ref{sec:timedep}.
In each model we kept the birth indices that come closest to our measurements at
$10^3$--$10^{3.5}$\,au, so these are predictions for data that we did not fit.
In the $10^{2.5}$--$10^3$\,au bin the timescale for dynamical evolution is so long (equation\cmark{~}\eqref{eq:ionize_numerical})
 that the model prediction is mostly reflecting the initial conditions.\footnote{
    Strictly speaking, our dynamical model should not be applied
 at $10^{2.5}$--$10^3$\,au anyway, because those binaries meet so few stars within
  a distance $\sim a$ during their lifetimes that
the diffusion approximation breaks down \citep{Hamilton2024,caputo2026evolution}.
 Nevertheless we show the results for illustrative purposes.}

From the top row of Figure~\ref{fig:seps} we see that
at $10^{2.5}$--$10^3$\,au, measured eccentricities
are close to the thermal distribution in every $\Delta G$
bin and every $J_z$ bin.
Evolution in the Galactic field is slower at these separations than at $10^3$--$10^{3.5}$\,au
\citep{Modak2023,Hamilton2024}, so a dependence of the birth distribution on mass ratio would still
be visible there at low $J_z$.
With the birth indices of $10^3$--$10^{3.5}$\,au, all three models predict such a dependence.
Their curves lie above most measurements in panels (a) and (b), and below the measurement in the lowest
$J_z$ bin of panel (c).
The birth eccentricity distribution at these separations is therefore close to thermal for
binaries of every mass ratio, except perhaps the twins.
We conclude that the birth eccentricity distribution depends on mass ratio only above about
$10^3$\,au, close to the separation at which \citet{hwang2022} found that the eccentricity
distribution changes from thermal to superthermal.
As for  $10^{3.5}$--$10^4$\,au (bottom row), the measurements there are less precise; they 
qualitatively repeat the trend 
that we found at $10^3$--$10^{3.5}$\,au, 
with superthermal eccentricities tending towards thermal as $J_z$ increases, but there are a couple of data points that
are scattered far from this expectation.
The curves of all three models pass through the error bar of the lowest $J_z$ bin in panel (d).
Only the curve of the model with a birth distribution that changes with time reaches the lowest $J_z$ bin in
panel (f).
In that model the low $\alpha$ of this bin comes from binaries younger than 1\,Gyr, which evolve
little at any separation.
In the other two models it comes from older binaries, which evolve faster at wider separations.
This bin holds only 72 binaries, so we do not use it to choose between the models.

%Evolution is faster at these separations, and faster early encounters bring the eccentricities
%of nearly all binaries close to thermal before we observe them.
%For $\Delta G<3$ and $\Delta G\geq7$ the curves for faster encounters therefore lie closer to
%$\alpha=1$ at $J_z<0.1\Jzu$ than our measurements, even with birth indices at the limits of our
%model.

Panels (a) and (d) of Figure~\ref{fig:seps} also show our measurements including the twins
($\Delta G<0.25$\,mag), as Figure~\ref{fig:fig2}(a) does for $10^3$--$10^{3.5}$\,au.
At all three separations the twins raise $\alpha$ most in one of the two lowest $J_z$ bins, and by
less than 0.1 in the two highest.
This pattern is consistent with twins being born at every epoch with an excess of very eccentric
orbits, of the kind that \citet{hwang2022b} found.
Dynamical evolution erases a narrow excess at $e\gtrsim0.95$ faster than a broad distribution, so
the effect of the twins on $\alpha$ fades with age and therefore with $J_z$.

\subsection{Open questions and future work}
\label{sec:origins}

Our observational results and their theoretical interpretation leave two major questions open.

The first is why the eccentricities of wide binaries at birth (or, more properly, while they are
dynamically young) depend so strongly on mass ratio (see Figure\cmark{~}\ref{fig:fig3}b).
A mixture of the formation channels of the Introduction whose relative importance depends on mass ratio could give binaries with
similar components more eccentric orbits than binaries with unequal components.
In simulations of star cluster formation, \citet{mathew2024wide} found that wide binaries with a
total mass below $0.8\,\mathrm{M}_\odot$ are much more superthermal than heavier ones.
This dependence on total mass cannot produce our trend with $\Delta G$, because the binaries in
our three $\Delta G$ bins have similar total masses.
For $J_z<0.1\Jzu$ the median total mass, which we estimated from absolute $G$ magnitudes as in
\S\ref{sec:sample}, lies between 1.2 and $1.5\,\mathrm{M}_\odot$ in each $\Delta G$ bin.
We do not know of a formation model that predicts the subthermal eccentricity distribution that we infer for binaries with low-mass companions, which makes these binaries an interesting target for models of wide-binary formation.

The second question is why this dependence is absent in older binaries ($\alpha\simeq1$ at the
largest $J_z$ in every $\Delta G$ bin of Figure~\ref{fig:fig2}).
Each of our two explanations has a cost, and we find neither satisfactory.
A boosted encounter rate (\S\ref{sec:faster}) must last for a few Gyr, which is longer than most birth
environments survive.
Such a boost also disrupts many old binaries, and to make up for these losses the birth rate must have fallen more
steeply than the measured one (Appendix~\ref{app:birth}).
Moreover, we imposed the boost by hand.
A proper physically-motivated model of a birth environment would instead follow the densities and relative
velocities that a binary meets while that environment disperses.
These differ between an unbound association that expands and a bound cluster that loses its stars gradually, and
a binary may leave its parent cluster long before the cluster dissolves.
Dispersal also binds some pairs of stars from the same cluster into new wide binaries
\citep{Kouwenhoven2010,Moeckel2011,Rozner2023}, so some wide binaries in the field today were not born as binaries.
A birth distribution that changes with time (\S\ref{sec:timedep}) instead requires that, among binaries formed in roughly the last Gyr, birth eccentricities depend on
mass ratio, whereas among older binaries they do not.
This could reflect a genuine change in how wide binaries formed over the history of the disc.
Alternatively, the binaries formed near the Sun in the last Gyr may come from a small number of
star-forming regions, whose conditions need not resemble those in which older binaries formed.
\citet{Hwang2021} found that the fraction of stars in wide binaries depends strongly on
metallicity, and that near solar metallicity it is higher for kinematically younger stars.
A boosted encounter rate would lower the fraction of old stars in wide binaries, and a formation environment
that changed with time could have changed the birth eccentricity distribution too.

Our interpretation rests on assumptions that future work should test.
Firstly, we assumed that the selection of our sample does not depend on eccentricity or on age, and tested
this assumption in \S\ref{sec:results}.
At low transverse velocity, however, the measured value of $\alpha$ depends on the radial-velocity requirement and
on the brightness of the primary (Appendix~\ref{app:vt}).
Whether this reflects differences in the underlying stellar populations sampled or magnitude-dependent observational
biases remains unclear. Binaries whose barycentres move slowly, which also have the smallest values of $J_z$, are therefore
the ones for which this assumption is least secure.
Secondly, we assumed that young binaries are heated vertically like other stars of their age.
If young binaries were born vertically colder than this, the lowest $J_z$ bin would contain more
of them, which would bring both explanations closer to the faint binaries in that bin.
Thirdly, although we account for the measurement uncertainties in $\gamma$ when inferring $\alpha$, the uncertainties reported in Table~\ref{tab:alpha_main} may not fully capture systematic errors in the \textit{Gaia} astrometric measurements. For this reason we have based our conclusions on differences in $\alpha$ that are large compared
with these uncertainties, rather than on the details of model fits.

Ages inferred from stellar parameters give the same trends as $J_z$ (Figure~\ref{fig:A_3}), so these
trends do not depend on using the $J_z$ of each binary as a proxy for its age.
These ages are uncertain by at least about 30 per cent, however, and cannot pick out binaries younger than 1\,Gyr. Two measurements would tell our explanations apart.
The first is $\alpha$ for binaries with precise individual ages.
Gyrochronology, isochrone ages of evolved primaries and membership of young moving groups can all
pick out binaries younger than about 1\,Gyr.
For such binaries we expect under both explanations that $\alpha$ depends on $\Delta G$ even more
strongly than in our lowest $J_z$ bin.
For binaries older than a few Gyr with $\Delta G\geq7$\,mag, we expect $\alpha<1$ if the encounter rate was
boosted and $\alpha>1$ if the birth distribution changed with time.
The second measurement is the fraction of stars that have a wide companion as a function of age,
which a boosted encounter rate would make fall steeply.
More precise measurements at $10^{3.5}$--$10^4$\,au would also help, because there the two
explanations predict different values of $\alpha$ for $\Delta G\geq7$ \,mag at low $J_z$
(Figure~\ref{fig:seps}f).
Finally, the shape of the eccentricity distribution is itself a test.
Our models predict the whole distribution of $\gamma$ in each bin, not only the index $\alpha$.
\citet{hwang2025white} found a possible lack of the most eccentric orbits among wide binaries with a white dwarf, which
they attribute to disruption.
A search for similar departures from a power law could help to distinguish explanations that disrupt many binaries from
those that do not. All of these tests will become sharper with \textit{Gaia} DR4, which will provide radial velocities and hence $J_z$ for many more binaries, and more precise
$v$--$r$ angles.

\section{Summary}
\label{sec:summary}

In this paper we measured how the eccentricity distribution of wide binaries depends on their
vertical action $J_z$ and on the magnitude contrast between their components $\Delta G$,
 and we interpreted the
results with simple models of their dynamical evolution.
Our results can be summarised as follows:
\begin{itemize}
    \item We measured the eccentricity index $\alpha$ of about 12\cmark{,}000 wide binaries with projected
    separations of $10^3$--$10^{3.5}$\,au (Section~\ref{sec:results}, Figure~\ref{fig:fig2},
    Table~\ref{tab:alpha_main}).
    At large $J_z$, binary eccentricities are roughly thermal ($\alpha \simeq 1$) regardless of
    $\Delta G$.
    At low $J_z$ they are much more sensitive to $\Delta G$, with binaries with the smallest (largest)
    $\Delta G$ having very superthermal (subthermal) eccentricities (Figure~\ref{fig:fig3}).

    \item We interpret these trends as a sign that binary eccentricities thermalize over time.
    We found qualitatively similar trends when we replaced $J_z$ by ages inferred from stellar parameters
    (Appendix~\ref{app:ages}, Figure~\ref{fig:A_3}).
    We also infer that wide binaries are born with eccentricities that depend strongly on mass ratio.
    Pairs of similar brightness are born very eccentric, and pairs with the faintest companions
    are born with low eccentricities.
    \item We modelled binaries born with an eccentricity distribution fixed in time, which then evolve under
    encounters with passing stars and the Galactic tide at the rates expected in the Galactic field today.
    In this fiducial model $\alpha$ evolves much too slowly to explain the trends that we observe
    (\S\ref{sec:model}, Figures~\ref{fig:fan} and\cmark{~}\ref{fig:const}).
    \item In our fiducial model we assumed a birth rate close to the measured history of star formation near
    the Sun, and with it we predicted numbers of binaries in each $J_z$ bin close to those that we observe
    (Appendix~\ref{app:birth}).
    When we changed the birth rate, the values of $\alpha$ that our models predict changed by less than our error
    bars.
    We regard encounters 10 to 100 times more frequent than in the Galactic field at all ages as implausible, because we could then match the observed number of binaries in each $J_z$ bin only with birth rates that fell far more
    steeply than measured.
    \item We found two models that could approximately explain the observations.
    In the first, we boosted the encounter rate far above that of the Galactic field for the first few Gyr of each binary's life (\S\ref{sec:faster}).
    We found that a boost lasting only for the first few hundred Myr, as we would expect of most birth environments,
    cannot explain the observations. In the second, we kept the encounter rate of our fiducial model but let only binaries born in roughly the last
    Gyr have eccentricities that depend on mass ratio (\S\ref{sec:timedep}).
    However, neither model seems physically well-motivated.
    \item At projected separations of $10^{3.5}$--$10^4$\,au the measurements are less precise but qualitatively repeat the trends of the 
    $10^3$--$10^{3.5}$\,au bin.
    By contrast, at $10^{2.5}$--$10^3$\,au we found no dependence
    of $\alpha$ on $\Delta G$ at any $J_z$, apart from the twins (\S\ref{sec:seps}, Figures~\ref{fig:fig3} and
    \ref{fig:seps}).
    We infer that binaries of every mass ratio are born with eccentricities close to thermal, and that the birth eccentricity distribution depends on mass ratio only above about $10^3$\,au.
\end{itemize}

% \section{Results}

% alpha vs Jz for 3-3.5 sep bin (in the appendix put it for the other two sep bin?)

% Counts per bin

% alpha vs delta G for the sep bin (put the other two sep bin in appendix)

% We can also put the inferred age one here as well, but explain the methodology in appendix?

% -- gaia
% -- super thermal
% -- tides encounters etc whether the eccentricity ev can be measured in the data
% -- gaia dr3
% -- vertical actions

% -- we find that tending towards thermal
% -- however there appears to be delta G
% -- initial conditions

% Outline:

% Original dataset (El Badry) - then Hwang's published catalogue with gamma angles - quality cuts - Then the radial velocity requirement and good RV cuts - final sample with and without twins - separations considered - any plots required? CMD? Mention WD contamination.

% Before/after RV selection gamma distribution may be? I think may be in the appendix or even in the main text; we should include one section where we discuss the selection effects, esp. introduced by the RV requirement. 

\section*{Acknowledgements}
We thank Andrea Caputo, Shaunak Modak, Adrian Price-Whelan, Sheila Sagear, and Giovanni Maria Tomaselli for helpful conversations. 
This work has made use of data from the European Space Agency (ESA) mission
{\it Gaia} (\url{https://www.cosmos.esa.int/gaia}), processed by the {\it Gaia}
Data Processing and Analysis Consortium (DPAC,
\url{https://www.cosmos.esa.int/web/gaia/dpac/consortium}). 
%Funding for the DPAC
%has been provided by national institutions, in particular the institutions
%participating in the {\it Gaia} Multilateral Agreement.
We used the large language models GPT--5.6 (OpenAI) and Claude (Anthropic) to help with code development and text editing. We checked their results and take full responsibility for the content of this paper.
%TODO (Chris, Lekshmi): the last two sentences of the Acknowledgements are a draft disclosure of the use
%TODO of AI tools, which MNRAS asks for; edit them to match what you both did, and use the same text in the
%TODO cover letter.
%TODO (Lekshmi): acknowledge LAMOST for the ages of Appendix A? Its standard wording: "Guoshoujing Telescope
%TODO (the Large Sky Area Multi-Object Fiber Spectroscopic Telescope LAMOST) is a National Major Scientific
%TODO Project built by the Chinese Academy of Sciences. Funding for the project has been provided by the
%TODO National Development and Reform Commission. LAMOST is operated and managed by the National Astronomical
%TODO Observatories, Chinese Academy of Sciences." (bib keys LAMOSTCui2012 and LAMOSTZhao2012 are in
%TODO example.bib.)
%%%%%%%%%%%%%%%%%%%%%%%%%%%%%%%%%%%%%%%%%%%%%%%%%%
\section*{Data Availability}

The wide-binary catalogues of \citet{elbadry2021} and \citet{hwang2022}, the {\it Gaia} DR3 data including the GSP-Spec stellar parameters and the age catalogue of \citet{Lu2026} are publicly available. The values of $\alpha$ in every bin, and the code of our dynamical model, are available from the
corresponding author on reasonable request.
%TODO (Lekshmi): is the Lu catalogue public? (Chris): "on reasonable request", or a Zenodo record for the
%TODO code and the tables of alpha?

%%%%%%%%%%%%%%%%%%%% REFERENCES %%%%%%%%%%%%%%%%%%

% The best way to enter references is to use BibTeX:

\bibliographystyle{mnras}
\bibliography{example} % if your bibtex file is called example.bib

% Alternatively you could enter them by hand, like this:
% This method is tedious and prone to error if you have lots of references
%\begin{thebibliography}{99}
%\bibitem[\protect\citeauthoryear{Author}{2012}]{Author2012}
%Author A.~N., 2013, Journal of Improbable Astronomy, 1, 1
%\bibitem[\protect\citeauthoryear{Others}{2013}]{Others2013}
%Others S., 2012, Journal of Interesting Stuff, 17, 198
%\end{thebibliography}

%%%%%%%%%%%%%%%%%%%%%%%%%%%%%%%%%%%%%%%%%%%%%%%%%%

%%%%%%%%%%%%%%%%% APPENDICES %%%%%%%%%%%%%%%%%%%%%
\appendix
\section{Tests of the inferred eccentricity indices}
\subsection{Observed and expected \texorpdfstring{$\gamma$}{gamma}  distributions}
\label{sec:obs_gamma_dist}

To assess whether the inferred values of $\alpha$ shown in Figure~\ref{fig:fig2} provide a good description of the observed $\gamma$ angle distributions, we forward model the expected $\gamma$ distributions for the corresponding three $\Delta G$ bins across the five $J_z$ bins in the separation range $3\leq \log_{10}(s/\mathrm{au})<3.5$. For each of these subsamples, we use the maximum a posteriori (MAP) value of $\alpha$ and follow the procedure described in Section~\ref{sec:inference} to obtain the corresponding $p(\gamma_{\rm true}\mid\alpha)$ distribution. To account for measurement uncertainties, we describe
$p(\gamma_{\rm obs}\mid\gamma_{\rm true},\sigma_{\gamma,i})$ for each binary $i$ using a Gaussian measurement kernel with width $\sigma_{\gamma,i}$, reflected at $0^\circ$ and $180^\circ$
and normalised over the range $0^\circ\leq\gamma_{\rm obs}\leq180^\circ$. We then marginalize over $\gamma_{\rm true}$ to obtain the expected observed distribution for each binary,
\begin{equation}
p_i(\gamma_{\rm obs}\mid \alpha)=\int_{0}^{180^\circ} p(\gamma_{\rm obs}\mid\gamma_{\rm true},\sigma_{\gamma,i})\,p(\gamma_{\rm true}\mid\alpha)\,{\rm d}\gamma_{\rm true}.
\end{equation}

Because the expected $\gamma$ distributions are symmetric about $90^\circ$, we fold the observed and predicted distributions about $90^\circ$. For each binary, we first calculate the probability that its observed $\gamma$ falls in each $15^\circ$ interval between $0^\circ$ and $180^\circ$. We then combine intervals symmetric about $90^\circ$, yielding six probabilities corresponding to folded $\gamma$ intervals between $0^\circ$ and $90^\circ$. For each of these six intervals, we sum the corresponding probabilities over all binaries in a given $\Delta G$--$J_z$ bin and divide by the total number of binaries to obtain the expected fraction. To quantify the sampling variation, we generate $10^4$ mock folded-$\gamma$ histograms for each $\Delta G$--$J_z$ bin. In each realization, we draw one of the six folded $\gamma$ intervals for every binary according to its individual set of six probabilities. The resulting set of draws forms one mock histogram with the same number of binaries as the corresponding observed $\Delta G$--$J_z$ bin. We then calculate the 16th--84th and 5th--95th percentiles of the fraction of binaries in each interval across all mock realizations.

\begin{figure*}
    \centering
    \includegraphics[width=0.95\linewidth]{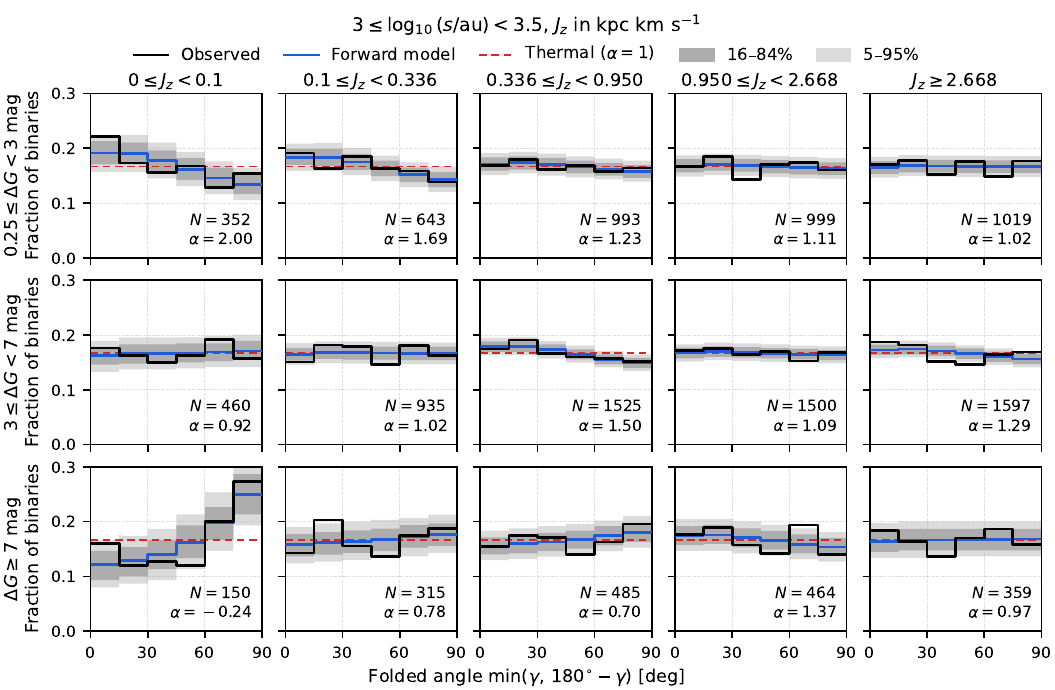}
    \caption{Folded distributions of the projected $v$--$r$ angle, $\min(\gamma,180^\circ-\gamma)$, for the
    binaries with $3\leq\log_{10}(s/\mathrm{au})<3.5$ without twins, in the 15 bins of $J_z$ and $\Delta G$ of
    Table~\ref{tab:alpha_main} ($J_z$ in kpc\,km\,s$^{-1}$).
    Black lines show the observed fraction of binaries in each $15^\circ$ bin.
    Blue lines show the fractions expected for $p(e)\propto e^{\alpha}$ with the MAP value of $\alpha$ given in
    each panel, including the uncertainty in $\gamma$ of each binary.
    Dark and light grey bands show the 16th--84th and 5th--95th percentiles of $10^4$ mock histograms drawn
    from the same expected distribution.
    Red dashed lines show the fraction expected for the thermal distribution, $\alpha=1$, for which $\gamma$ is
    distributed uniformly, so that each bin holds one sixth of the binaries.
    Each panel also gives the number of binaries $N$.}
    \label{fig:C_1}
\end{figure*}

Figure~\ref{fig:C_1} summarises the results, and we described its trends in Section~\ref{sec:results}.
In every bin of $\Delta G$ and $J_z$, the observed distribution follows the expected one.
Only 6 of the 90 observed fractions lie outside the 5th--95th percentile band of the mock histograms, about as many as
expected by chance.
The inferred values of $\alpha$ therefore describe the observed distributions of $\gamma$.

\subsection{What \texorpdfstring{$\alpha$}{alpha} measures}
\label{app:alpha}

To be clear, we use the power law only to summarise the distribution of $\gamma$ in a population
by a single number, and not because we expect the eccentricities of wide binaries to follow an exact power law.
For a thermal distribution ($\alpha=1$) the direction of the relative velocity is isotropic and
independent of the separation vector, so $\gamma$ is distributed uniformly between $0^\circ$ and
$180^\circ$.
A population with more eccentric orbits than this ($\alpha>1$) has an excess of binaries near
$\gamma=0^\circ$ and $180^\circ$, and a population with less eccentric orbits ($\alpha<1$) has an
excess near $90^\circ$.
We can quantify this excess by the mean of $\cos2\gamma$ over the binaries in a population,
$\langle\cos2\gamma\rangle$, which is zero for a thermal distribution and positive for $\alpha>1$.
Most of the information about $\alpha$ in the angles lies in $\langle\cos2\gamma\rangle$.
To see why, note that the distribution of $\gamma$ is symmetric about $90^\circ$, because a binary
spends as long moving towards pericentre as moving away from it, and reversing the relative
velocity changes $\gamma$ into $180^\circ-\gamma$.
The distribution can therefore be written as a sum of the harmonics $\cos2\gamma$, $\cos4\gamma$,
$\cos6\gamma$ and so on, of which $\cos2\gamma$ is the broadest.
Each eccentricity gives its own pattern of peaks in $\gamma$.
In a population with a broad range of eccentricities these peaks blur into a smooth excess at the
ends or in the middle of the range of $\gamma$, and we can describe such an excess by the harmonic
$\cos2\gamma$ alone.
The exception is a population with many nearly circular orbits, all of which add to a narrow peak
at $90^\circ$.
Using the grid of \citet{hwang2022}, we found that an estimate of $\alpha$ from
$\langle\cos2\gamma\rangle$ alone has an uncertainty only 2 per cent larger than the full
posterior when $\alpha=1$.
When $\alpha=0$, nearly circular orbits are more common and the uncertainty is 10 per cent larger.

We checked how well a power law summarises our measurements in the 15 bins of $J_z$ and $\Delta G$
at $10^3$--$10^{3.5}$\,au of Figure~\ref{fig:C_1}.
In every bin, the value of $\alpha$ that matches $\langle\cos2\gamma\rangle$ lies within 0.15 of our
MAP estimate.
We also resampled the binaries in each bin with replacement, which does not assume that the
eccentricities follow a power law.
The resulting spread in the MAP estimate is within about 10 per cent of the width of our 68 per cent HDI.
Unfortunately, a few hundred binaries are not enough to tell a power law from many other
eccentricity distributions with similar $\langle\cos2\gamma\rangle$, such as an equal mixture of
two power laws with $\alpha=3$ and $\alpha=-0.5$.
We therefore read each value of $\alpha$ as a measure of how strongly a population favours high or
low eccentricities. For example, a population in which every binary has $e\simeq0.25$ has nearly the same
distribution of $v$--$r$ angles as $p(e)\propto e^{\alpha}$ with $\alpha\simeq-0.8$.
Telling these two apart would take several hundred binaries, whereas the lowest $J_z$ bin for
$\Delta G\geq7$ at $10^3$--$10^{3.5}$\,au contains 150.
When we infer in \S\ref{sec:timedep} that young binaries with faint companions are born with low
eccentricities, we therefore mean only that few of them are born with high eccentricities.

\section{Testing the observed trends using alternative age diagnostics}
\subsection{$\alpha$ as a function of inferred ages}
\label{app:ages}

As a test of our $\alpha$ versus $J_z$ trends across different $\Delta G$ bins, we examine the same trends as a function of stellar age. Of course, stellar ages are hard to obtain precisely, and age uncertainties could smooth out any underlying trends. One could potentially infer ages from a binary's $J_z$ or $\sigma_z$ using existing kinematic age calibrations in the literature. However, this would not provide an independent check of the $\alpha (J_z,\Delta G)$ trends we reported in Section~\ref{sec:results}. Our goal here is not to determine precise ages for individual binaries, but rather to order them by relative age without relying on their own $J_z$ measurements.

We infer ages using the stellar age catalogue of \cite{Lu2026} (hereafter the Lu catalogue), which contains ensemble kinematic ages for 1.5 million LAMOST stars. Their catalogue is built by calibrating an empirical relation between stellar age, median $\ln J_z$, and [Fe/H] using the precise isochrone ages of subgiant stars from \cite{xiang2022}. \cite{Lu2026} then apply this calibration to the LAMOST stars by identifying an ensemble of stars in a 13-D stellar parameter space including $T_{\mathrm{eff}}$, $\log g$, [Fe/H], and ten elemental abundances, for each LAMOST star. The ensemble stars are drawn from a subset of the LAMOST stellar sample with full 6D phase-space measurements available from \textit{Gaia} DR3 to calculate $J_z$. They then use the median $\ln J_z$ of the ensemble stars to infer an ensemble age from the calibrated relation. They report that typical age uncertainties in their catalogue are of the order of 30 per cent. We further restrict the Lu catalogue to stars with ensemble ages $\leq10$ Gyr, as \cite{Lu2026} report that their inferred ages start to flatten around this threshold.

\begin{figure}
    \centering
    \includegraphics[width=0.95\columnwidth]{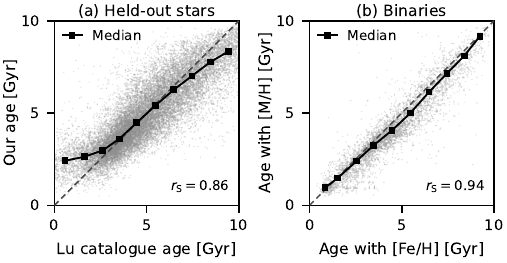}
    \caption{Validation of the stellar age inference. \textit{(a)} Ages inferred using the median age of the
    30 nearest neighbours in the standardized 3D parameter space defined by $T_{\rm eff}$, $\log g$, and
    [Fe/H], compared with the ensemble ages reported by \citet{Lu2026} for a randomly selected validation sample
    of 20,000 stars. \textit{(b)} Comparison of ages inferred for wide binaries using [Fe/H] and [M/H] in the
    3D nearest-neighbour method. Grey points show individual stars or binaries, black squares joined by a line
    show the median relation in 1-Gyr bins, and the dashed line indicates the 1:1 relation.
    $r_{\rm S}$ is the Spearman rank correlation coefficient.}
    \label{fig:A_1_2}
\end{figure}

 \begin{figure*}
    \centering
    \includegraphics[width=0.9\linewidth]{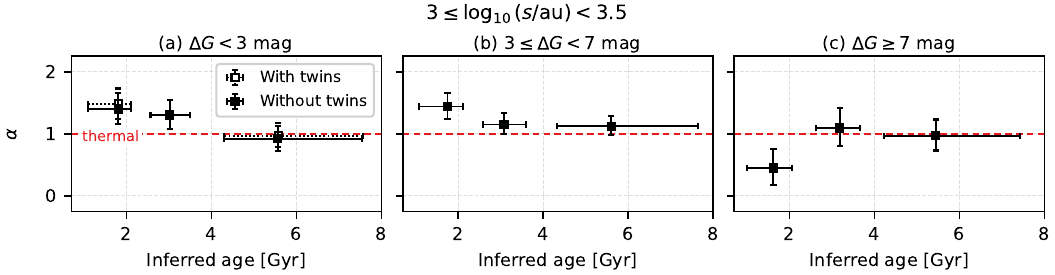}
    \caption{As Figure~\ref{fig:fig2}, with the inferred stellar age in place of $J_z$.}
    \label{fig:A_3}
\end{figure*}
We cross-match the wide binaries in the separation range $3\leq\log_{10}(s/\mathrm{au})<3.5$ with the \textit{Gaia} DR3 astrophysical parameter catalogue to obtain $T_{\mathrm{eff}}$, $\log g$, and [M/H] derived from the GSP-Spec pipeline. Of the 12,234 binaries in this separation range, 9,055 have all three stellar parameters available for at least one component. If both binary components have these measurements available, we use the brighter component. Otherwise, we use the component with available measurements. Since \textit{Gaia} DR3 does not provide all the elemental abundances used in \cite{Lu2026}, we instead use a 3D stellar parameter space consisting of $T_{\mathrm{eff}}$, $\log g$, and metallicity. While the Lu catalogue uses [Fe/H] for stellar parameter matching, requiring [Fe/H] measurements reduces our binary sample size to roughly half. Therefore, to retain statistics, we use [Fe/H] $\simeq$ [M/H] while inferring ages from the Lu catalogue.

We standardize the stellar parameters using \texttt{StandardScaler} from \texttt{sklearn.preprocessing} fitted to the Lu catalogue, and apply the same transformation to the stellar parameters of the wide binary catalogue. Then for each binary, we find the 30 nearest neighbours in the standardized 3D parameter space defined by $T_{\mathrm{eff}}$, $\log g$, and [Fe/H] using the \texttt{cKDTree} algorithm from \texttt{scipy.spatial}, and assign the median ensemble age of neighbours.

To test whether inferring ages in the 3D parameter space preserves the age ranking, we randomly select 20,000 stars from the Lu catalogue and remove them from the catalogue. 
We repeat the standardization based on the remaining stars and for each star, we find the 30 nearest neighbours among the remaining stars in the Lu catalogue in the standardized 3D parameter space defined by $T_{\mathrm{eff}}$, $\log g$, and [Fe/H]. We assign the median age of these 30 stars and compare the recovered ages against their original ensemble ages. Figure\cmark{~}\ref{fig:A_1_2}a summarises the results. We find a strong correlation between the two age estimates with Spearman's rank correlation coefficient of 0.860. Comparing the median trend of the 3D ensemble ages in 1 Gyr bins of the Lu catalogue ages with the 1:1 line shows that the ages of younger stars are overestimated, whereas those of older stars are underestimated. However, the relative age ordering is broadly preserved.

\begin{table}
\centering
\small
\setlength{\tabcolsep}{6pt}
\caption{Numbers of binaries with $3\leq\log_{10}(s/\mathrm{au})<3.5$ in the bins of inferred age and
$\Delta G$ of Figure~\ref{fig:A_3}.
Numbers in brackets exclude the twins ($\Delta G<0.25$\,mag).}
\label{tab:alpha_age}
\begin{tabular}{lccc}
\hline
& \multicolumn{3}{c}{$\Delta G$ [mag]} \\
Inferred age [Gyr] & $0$--$3$ & $3$--$7$ & $\geq7$ \\
\hline
$0$--$2.32$ & 910 (812) & 1,314 & 321 \\
$2.32$--$3.90$ & 918 (830) & 1,561 & 503 \\
$\geq3.90$ & 915 (837) & 1,929 & 684 \\
\hline
\end{tabular}

\end{table}
To understand the impact of the approximation [Fe/H] $\simeq$ [M/H] on the age inference, we select the binaries with both [Fe/H] and [M/H] available, and infer their ages separately using [Fe/H] and [M/H]. The results are summarised in Figure~\ref{fig:A_1_2}b. The Spearman correlation coefficient between the two age estimates is 0.941, suggesting that the relative age ordering is broadly preserved. However, the median trend relative to the 1:1 line indicates that the approximation tends to underestimate ages, especially at intermediate ages.

Having found that the method broadly recovers the age ranking
in the reference catalogue, we now examine how $\alpha$ varies
with inferred age. We divide the wide binary sample into three different $\Delta G$  bins: 0--3, 3--7 and $\geq7$ mag. We then divide the $\Delta G=0$--3 sample into three equal count bins based on the inferred binary ages. We then apply the same age bin boundaries to the latter two $\Delta G$ bins and infer $\alpha$ as a function of inferred ages. We repeat the analysis by excluding binaries with $\Delta G<0.25$, corresponding to the twins. The results are summarised in Figure~\ref{fig:A_3}, and Table~\ref{tab:alpha_age} gives the numbers of binaries in each bin. Figure\cmark{~}\ref{fig:A_3} qualitatively reproduces the trends seen in Figure\cmark{~}\ref{fig:fig2}. In the lowest $\Delta G$ bin, the eccentricity distribution transitions from superthermal towards thermal with increasing inferred age, whereas in the highest $\Delta G$ bin, it tends to evolve from subthermal towards thermal. This is consistent with the expected dynamical evolution towards a thermal distribution while also suggesting that the dependence of $\alpha$ on $\Delta G$ in younger populations is real. Although the age bins are relatively broad, the typical age uncertainties of $\sim30$ per cent in the Lu catalogue, along with additional uncertainties introduced by the age inference in the 3D stellar parameter space, are likely to smooth out the intrinsic variation of $\alpha$ with age. Therefore, the observed age dependence of $\alpha$ may be an underestimate of the underlying trend.

\subsection{Transverse velocity as a proxy for ages}
\label{app:vt}
\begin{figure*}
    \centering
    \includegraphics[width=0.9\linewidth]{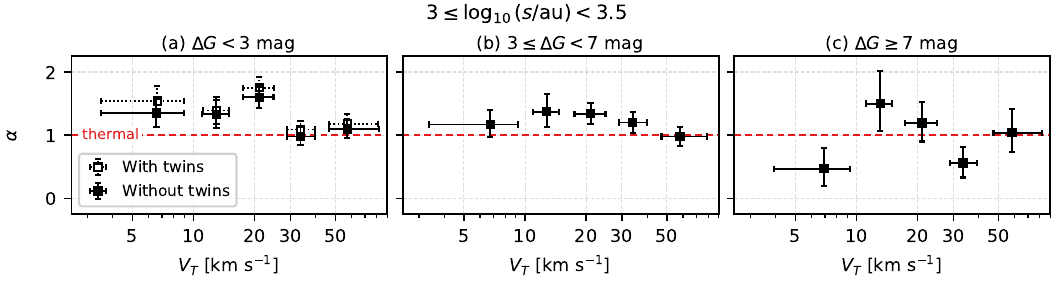}
     \caption{As Figure~\ref{fig:fig2}, with the transverse peculiar velocity $V_T$
    (Appendix~\ref{app:vt}) in place of $J_z$, for all binaries with or without a radial velocity.
    }
    \label{fig:B_1}
\end{figure*}

\begin{figure*}
    \centering
    \includegraphics[width=0.855\linewidth]{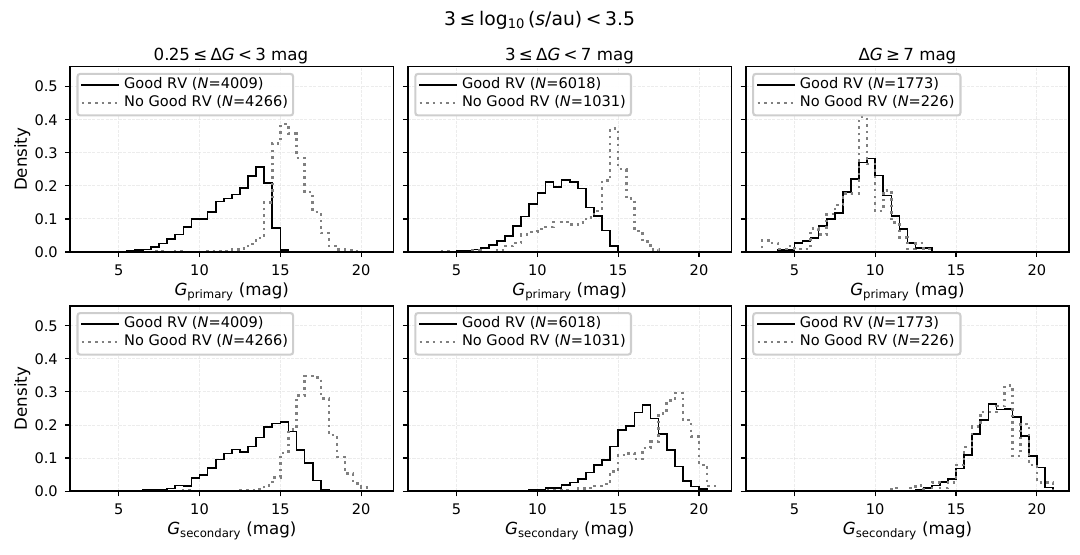}
    \caption{Distributions of the apparent \textit{Gaia} $G$ magnitudes of the primaries (top row) and the secondaries
    (bottom row) of the binaries with $3\leq\log_{10}(s/\mathrm{au})<3.5$, in three bins of $\Delta G$. Black solid lines show the binaries with good radial velocity (the Good RV sample), and grey dotted lines show the binaries without one (the No Good RV sample). The Good RV sample includes the retrograde binaries excluded from the main analysis.
    Each histogram is normalised to unit area, and the legends give the number of binaries in each sample.}
    \label{fig:B_2}
\end{figure*}

\begin{table}
\centering
\small
\setlength{\tabcolsep}{6pt}
\caption{Numbers of binaries with $3\leq\log_{10}(s/\mathrm{au})<3.5$ in the bins of transverse velocity
$V_T$ and $\Delta G$ of Figure~\ref{fig:B_1}.
Numbers in brackets exclude the twins ($\Delta G<0.25$\,mag).}
\label{tab:alpha_vt}
\begin{tabular}{lccc}
\hline
& \multicolumn{3}{c}{$\Delta G$ [mag]} \\
$V_T$ [km\,s$^{-1}$] & $0$--$3$ & $3$--$7$ & $\geq7$ \\
\hline
$0$--$10.14$ & 1,181 (1,001) & 931 & 290 \\
$10.14$--$15.78$ & 1,181 (1,038) & 895 & 280 \\
$15.78$--$26.70$ & 2,362 (2,032) & 1,731 & 505 \\
$26.70$--$43.03$ & 2,362 (2,076) & 1,770 & 522 \\
$>43.03$ & 2,362 (2,128) & 1,722 & 402 \\
\hline
\end{tabular}

\end{table}

We also study the variation of $\alpha$ as a function of the magnitude of the 2D projection of the total peculiar velocity vector onto the plane of the sky ($V_T$). We calculate $V_T$ using the parallax and proper motions of the brighter component of a given binary. We correct for solar motion and local differential rotation using $(U_\odot,V_\odot,W_\odot)=(11.1,12.24,7.25)\,\mathrm{km\,s^{-1}}$ \citep{schonrich2010} and the Oort constants $(A,B,C,K)=(15.3,-11.9,-3.2,-3.3)\,\mathrm{km\,s^{-1}\,kpc^{-1}}$ \citep{bovy2017}, respectively. Although $V_T$ is a weaker statistical proxy for ages than $J_z$, it lets us use the parent wide binary sample consisting of 78,013 candidates for the analysis, as calculating $V_T$ does not require radial velocities (RV). We generally expect the $\alpha$ versus $V_T$ trend to be more noisy than the $\alpha$ versus $J_z$ trend. 

Figure\cmark{~}\ref{fig:B_1} shows the $\alpha$ versus $V_T$ trend with and without twin binaries in different $\Delta G$ bins, for the separation range $3\leq\log_{10}(s/\mathrm{au})<3.5$. We define the $V_T$ bins by dividing the $\Delta G=0$--3 mag range into four equal count bins and then further dividing the low $V_T$ bin at its median to isolate the kinematically coldest population. We then apply the same $V_T$ boundaries to the other $\Delta G$ bins. Table~\ref{tab:alpha_vt} gives the numbers of binaries in each bin.
While the $\alpha$ versus $V_T$ trend within each $\Delta G$ panel is non-monotonic and noisy in general, the lowest $V_T$ bin shows a monotonic decline in $\alpha$ across different $\Delta G$, similar to what we observe in the lowest $J_z$ bin. The recovery of this trend without imposing an RV requirement is encouraging. However, to understand the impact of this selection on our results, we further investigate how the RV requirement affects the inferred eccentricity distribution. 

\subsubsection{Sensitivity of inferred $\alpha$ to RV selection}
\label{sec:appendix_radvel}

To understand whether the RV selection could contribute to the $\alpha$ versus $\Delta G$ trend reported in Figures\cmark{~}\ref{fig:fig2} and \ref{fig:fig3}, we compare the eccentricity distribution of wide binaries included in and excluded from our main analysis as a function of $V_T$. While the absence of such differences would not rule out selection effects in the $J_z$ analysis, their presence would warrant caution in interpreting our results.

\begin{figure}
    \centering
    \includegraphics[width=\columnwidth]{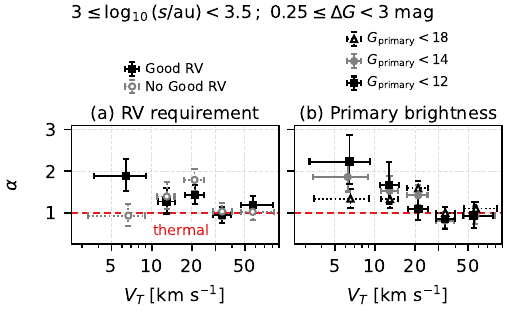}
    \caption{Eccentricity index $\alpha$ against transverse velocity $V_T$ for binaries with
    $3\leq\log_{10}(s/\mathrm{au})<3.5$ and $0.25\leq\Delta G<3$\,mag, in the $V_T$ bins of
    Figure~\ref{fig:B_1}.
    (a) Binaries with a good radial velocity (black filled squares) and without one (grey open circles).
    (b) Binaries of both kinds whose primaries are brighter than $G=18$ (black open triangles with dotted bars), 14 (grey filled circles) and 12 (black filled squares).}
    \label{fig:B_3}
\end{figure}

\textit{(i) The RV requirement changes the photometric composition of the sample.} Figure\cmark{~}\ref{fig:B_2} shows the distribution of the apparent \textit{Gaia} $G$-band magnitudes of the primary (top row) and the secondary (bottom row) components of the binaries in the RV-selected sample (hereafter, the Good RV sample) and those excluded by the RV requirement (hereafter, the No Good RV sample), across three different $\Delta G$ bins. Since brighter primaries are more likely to have reliable radial velocity measurements, the RV requirement preferentially removes binaries with relatively fainter primaries in the first two $\Delta G$ bins, with a corresponding shift towards brighter secondaries. The fraction of binaries excluded by the RV requirement also varies considerably with $\Delta G$. In the $0.25\leq\Delta G<3$ mag bin, 4,266 out of 8,275 binaries are excluded, compared to 1,031 out of 7,049 in the $3\leq\Delta G<7$ mag bin and 226 out of 1,999 in the $\Delta G\geq7$ mag bin. This trend is consistent with the correlation between $\Delta G$ and primary apparent magnitude in the parent catalogue. Since the catalogue is limited by the detectability of the fainter secondary components, binaries with larger $\Delta G$ must have brighter primaries to be included \citep{elbadry2021}.

\textit{(ii) The RV requirement changes the inferred eccentricity distribution.} Since the $0.25\leq\Delta G<3$ mag bin contains comparable numbers of Good RV and No Good RV binaries, providing sufficient statistics for both populations, we now focus on this bin to investigate whether the inferred eccentricity distribution is sensitive to the radial velocity requirement. Figure\cmark{~}\ref{fig:B_3}a compares $\alpha$ as a function of $V_T$ for the two samples, using the same $V_T$ boundaries as in Figure\cmark{~}\ref{fig:B_1}.

We find that the inferred eccentricity distribution in the lowest $V_T$ bin is strongly sensitive to the radial velocity requirement. The Good RV sample is strongly superthermal for $V_T\lesssim10$ \,km\,s$^{-1}$, with $\alpha=1.89_{-0.37}^{+0.41}$, comparable to the low $J_z$ bin in our main analysis. In contrast, the No Good RV sample is close to a thermal distribution, with $\alpha=0.93_{-0.26}^{+0.27}$. Interestingly, the difference between the two samples is not uniform across the $V_T$ bins; at higher $V_T$, the differences are less pronounced, and the 68 per cent HDIs overlap. This raises the question of why the inferred eccentricity distribution in the lowest $V_T$ bin is particularly sensitive to the RV requirement. 

Although $V_T$ and $J_z$ are expected to be statistically correlated as they both trace dynamical heating, low $V_T$ does not necessarily imply low $J_z$, and the lowest bins need not contain the same binaries. Nevertheless, this sensitivity to RV selection is relevant to our finding that the strongly superthermal eccentricity distribution in the lowest $J_z$ bin cannot be explained by our fiducial model of dynamical evolution assuming the same birth eccentricity distribution as the higher $J_z$ populations (see Figure~\ref{fig:fan}). Since the RV requirement preferentially selects binaries with brighter primaries, the observed difference in $\alpha$ could reflect differences in the underlying stellar populations sampled. We therefore investigate whether progressively excluding fainter primaries from the full sample, without imposing any RV requirement, shifts the inferred $\alpha$ towards that of the Good RV sample.

% This leads to an interesting possibility tied to binary formation: the dynamically colder population may contain binaries formed through different formation channels, with different birth eccentricity distributions. If selecting binaries based on primary brightness alters the relative contributions of these populations, imposing an RV requirement could change their relative fractions and hence the inferred $\alpha$.

\textit{(iii) The inferred eccentricity distribution appears sensitive to primary brightness in the lowest $V_T$ bin.} We consider three apparent magnitude limits, $G_{\mathrm{primary}}<18$, $<14$, and $<12$ mag, and infer $\alpha$ as a function of $V_T$ for the full sample using the same bin boundaries as before. The results are shown in Figure\cmark{~}\ref{fig:B_3}b. The lowest $V_T$ bin shows the largest change in the MAP estimates of $\alpha$ as we progressively select brighter primaries, with the inferred eccentricity distribution shifting towards increasingly superthermal values. $\alpha$ increases from $1.34_{-0.22}^{+0.23}$ for $G_{\mathrm{primary}}<18$ to $1.86_{-0.35}^{+0.41}$ for $G_{\mathrm{primary}}<14$ and $2.23_{-0.54}^{+0.63}$ for $G_{\mathrm{primary}}<12$. The estimate for $G_{\mathrm{primary}}<14$ is comparable to that of the Good RV sample. Note that since the magnitude-selected samples overlap, the inferred $\alpha$ estimates have correlated uncertainties.

The apparent magnitude depends on both the distance and the absolute magnitude of a star, so the differences in $\alpha$ estimates for different $G_{\mathrm{primary}}$ thresholds could reflect changes in both. We find that the distance distributions remain relatively similar, with the median distance decreasing only from 149 to 137 pc between the $G_{\mathrm{primary}}<18$ and $<12$ samples, whereas the median absolute magnitude changes considerably from $M_G=8.44$ to $4.51$ mag. Therefore, the observed change in $\alpha$ may be associated with differences in intrinsic luminosity, which could reflect differences in the underlying stellar populations, potentially including their ages and birth environments. If the lowest $V_T$ bin contains binaries formed through different formation channels, with different birth eccentricity distributions, selecting brighter primaries could alter their relative fractions and hence the inferred $\alpha$. However, we cannot rule out magnitude-dependent observational biases that may preferentially select binaries at certain orbital configurations and thereby affect the inferred eccentricity distribution. 

Since binaries with larger $\Delta G$ preferentially have brighter primaries in the parent catalogue, this raises the question of whether the observed $\alpha$--$\Delta G$ dependence is associated with $\Delta G$ itself or instead reflects the accompanying variation in primary brightness. Larger $\Delta G$ preferentially selects brighter primaries, whereas within the $0.25\leq\Delta G<3$ mag bin, progressively selecting brighter primaries in the lowest-$V_T$ population shifts $\alpha$ towards more superthermal values. These trends therefore act in opposite directions, suggesting that the decline of $\alpha$ with increasing $\Delta G$ is unlikely to arise solely from the associated variation in primary brightness. However, this is only suggestive, as the comparison is based on the lowest-$V_T$ population rather than the lowest-$J_z$ population.

\section{Additional tests of the observed trends}

\begin{figure*}
    \centering
    \includegraphics[width=0.92\linewidth]{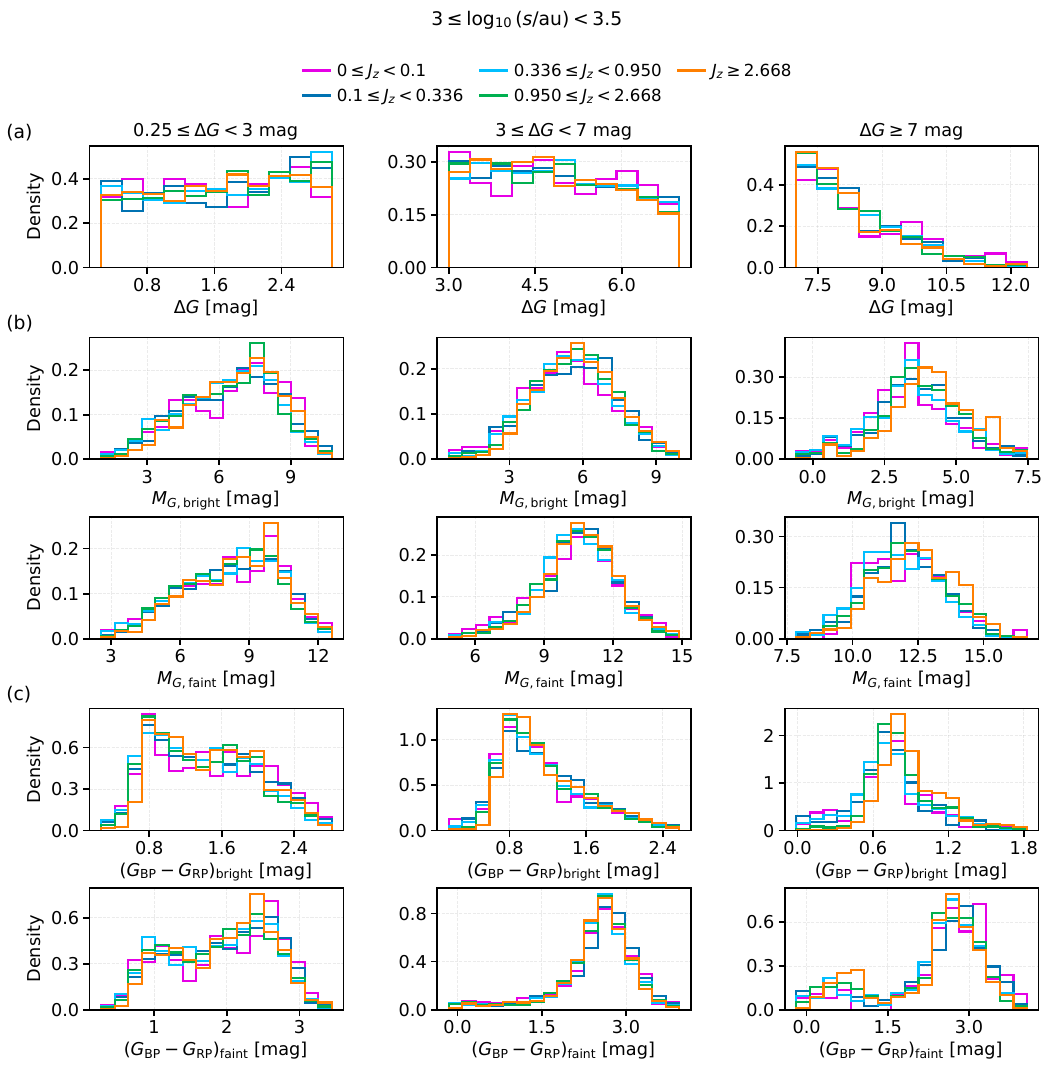}
    \caption{Distributions of the binaries with $3\leq\log_{10}(s/\mathrm{au})<3.5$, without twins, in the five
    $J_z$ bins of Figure~\ref{fig:fig2} (colours, $J_z$ in kpc\,km\,s$^{-1}$) and three bins of $\Delta G$
    (columns).
    (a) $\Delta G$.
    (b) Absolute $G$ magnitudes of the brighter (upper row) and fainter (lower row) components.
    (c) $G_\mathrm{BP}-G_\mathrm{RP}$ colours of the brighter (upper row) and fainter (lower row) components.
    Each histogram is normalised to unit area.}
    \label{fig:C2_3}
\end{figure*}

\begin{figure*}
    \centering
    \includegraphics[width=0.9\linewidth]{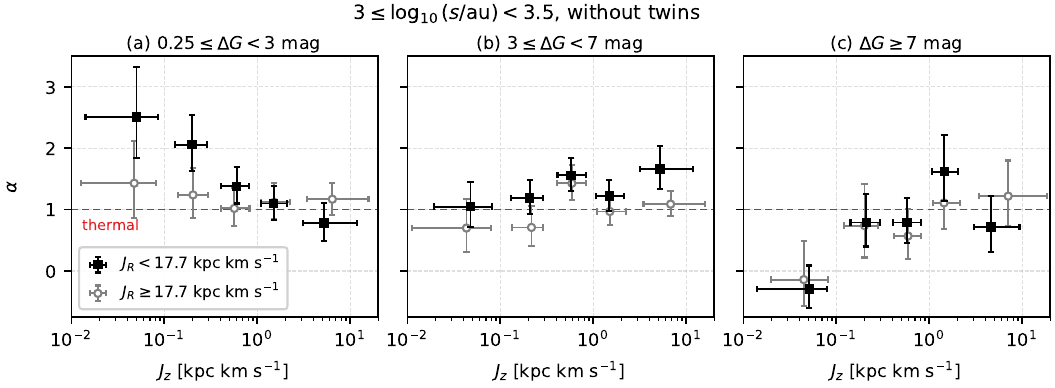}
    \caption{As Figure~\ref{fig:fig2} without the twins, split by radial action $J_R$ at the median $J_R$ of
    the binaries with $0.25\leq\Delta G<3$\,mag.
    Filled black squares show binaries with $J_R<17.7$\,kpc\,km\,s$^{-1}$ and open grey circles binaries with
    larger $J_R$.}
    \label{fig:C2_4}
\end{figure*}

\begin{figure}
    \centering
    \includegraphics[width=0.9\columnwidth]{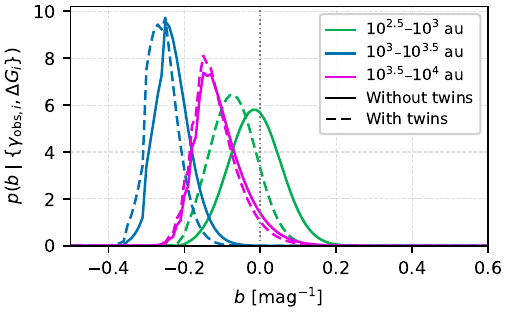}
    \caption{Posterior distributions of the slope $b$ of $\alpha=a+b\Delta G$ fitted to the individual binaries
    with $J_z<0.1$\,kpc\,km\,s$^{-1}$, in three ranges of projected separation (colours).
    Solid lines are for the binaries without twins and dashed lines for the binaries with twins.
    The dotted vertical line marks $b=0$.}
    \label{fig:C2_2}
\end{figure}

\begin{figure}
    \centering
    \includegraphics[width=0.9\columnwidth]{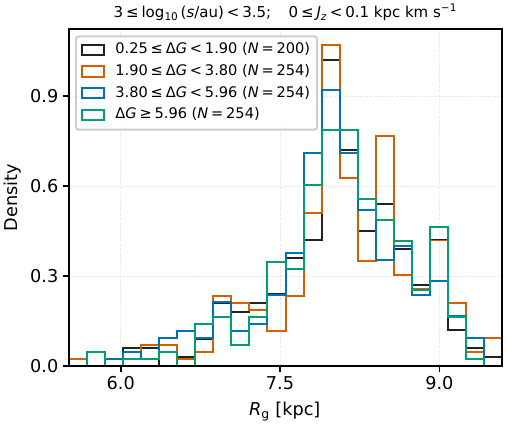}
    \caption{Distributions of the guiding radius $R_\mathrm{g}$ of the binaries with
    $3\leq\log_{10}(s/\mathrm{au})<3.5$ and $J_z<0.1$\,kpc\,km\,s$^{-1}$, without twins, in the four $\Delta G$
    bins of Figure~\ref{fig:fig3}.
    Each histogram is normalised to unit area, and the legend gives the number of binaries in each bin.}
    %TODO (Lekshmi): the figure title says 'AU' and 'kpc km/s'; house style is au and kpc km s^-1.
    % DONE
    \label{fig:C2_1}
\end{figure}

% \begin{figure*}
%     \centering
%     \includegraphics[width=\linewidth]{deltaG_distributions_across_Jz.png}
%     \caption{}
%     \label{fig:C2_3}
% \end{figure*}

\subsection{The $\alpha$--$J_z$ trend within the $\Delta G$ bins}
\label{app:jz}
(i) \textit{Photometric distributions across $J_z$ bins:} Since $\alpha$ depends on $\Delta G$, we test whether the observed trend with $J_z$ could instead arise from differences in the underlying $\Delta G$ distributions. Figure\cmark{~}\ref{fig:C2_3}a shows the $\Delta G$ distributions of binaries across the different $J_z$ bins, which show no obvious systematic differences. We also compare the absolute-magnitude distributions of the brighter and fainter components, as well as their colour distributions, across the different $J_z$ and $\Delta G$ bins, as shown in Figures\cmark{~}\ref{fig:C2_3}b and \ref{fig:C2_3}c. As discussed earlier in Section\cmark{~}\ref{sec:appendix_radvel}, increasing $\Delta G$ selects progressively brighter primaries and fainter secondaries, which is also reflected in their colour distributions. Within a given $\Delta G$ bin, the absolute-magnitude and colour distributions are broadly similar across the different $J_z$ bins. In the $\Delta G\geq7$ mag bin, the medians of the primary absolute magnitude and colour shift slightly in the highest-$J_z$ bin, but the distributions overlap substantially. Overall, we find no clear variation in the photometric distributions with $J_z$ that mirrors the observed variation in $\alpha$.

(ii) \textit{Split by radial action ($J_R$)}: Binaries in a given $J_z$ bin can have different radial excursions in the Galactic disc, and therefore span a range of $J_R$. We divide the binary population in the $0.25\leq\Delta G<3$ bin at the median $J_R$ and use the same threshold for the other two $\Delta G$ bins. We then examine how $\alpha$ varies with $J_z$ in the resulting low- and high-$J_R$ populations. The results are summarised in Figure~\ref{fig:C2_4}. In the $0.25\leq\Delta G<3$ bin, the low $J_R$ subsample shows a relatively steeper decline from superthermal to thermal distribution as a function of $J_z$, whereas the higher $J_R$ population is compatible with a flatter trend. Such a clear decline is not apparent in the other $\Delta G$ bins. However, the decrease in $\alpha$ with increasing $\Delta G$ in the lowest-$J_z$ bin is present in both radial-action subsamples. The steeper $\alpha$ versus $J_z$ trend in the lowest $\Delta G$ bin for the low-$J_R$ population may be because selecting binaries with both low $J_R$ and low $J_z$ isolates a dynamically younger population, whose eccentricity distribution may be closer to its birth eccentricity distribution.

\subsection{$\alpha$--$\Delta G$ trend within the lowest $J_z$ bin}
\label{app:dG}

(i) \textit{Sensitivity to binning in $\Delta G$:} To test whether the inferred dependence of the eccentricity distribution on $\Delta G$ is sensitive to our choice of $\Delta G$ bins, we also extend the Bayesian framework described in equation~\eqref{eq:posterior} to infer the dependence of the eccentricity distribution on $\Delta G$ directly instead of binning by $\Delta G$ as an additional test. Assuming a linear model $\alpha_i=a+b\Delta G_i>-1$, for each binary, where $a$ and $b$ are population-level parameters, we infer the joint posterior probability distribution $p(a,b\mid\{\gamma_{\mathrm{obs},i},\Delta G_i\})$ for each separation bin, given by
\begin{align}
&p(a,b\mid\{\gamma_{\mathrm{obs},i},\Delta G_i\})
\notag\\
&\propto p(a,b)\prod_{i=1}^{N}
\int_0^1\int_0^{180^\circ}
p(\gamma_{\mathrm{obs},i}\mid
\gamma_{\mathrm{true},i},\sigma_{\gamma,i})
p(\gamma_{\mathrm{true},i}\mid e_i)
\notag\\
&\hphantom{\propto p(a,b)\prod_{i=1}^{N}}
\times p(e_i\mid a,b,\Delta G_i)
\,d\gamma_{\mathrm{true},i}\,de_i.
\end{align}

We adopt a uniform joint prior on $a$ and $b$ over the ranges $[-0.99,5]$ and $[-1,1]$, respectively, retaining only combinations that satisfy the constraint $a+b\Delta G_i>-1$ for every binary in the sample. We evaluate the posterior on a grid with spacings $\Delta a=0.1$ and $\Delta b=0.01$. We then marginalize over $a$ to obtain $p(b\mid\{\gamma_{\mathrm{obs},i},\Delta G_i\})$ and calculate the posterior probability of a negative slope under the assumed linear model as $P(b<0\mid\{\gamma_{\mathrm{obs},i},\Delta G_i\})=\int_{-1}^{0}p(b\mid\{\gamma_{\mathrm{obs},i},\Delta G_i\})\ db$. The results are summarised in Figure~\ref{fig:C2_2}.

For binaries in the separation range $3\leq\log_{10}(s/\mathrm{au})<3.5$, the marginal posterior favours a negative slope, with $b=-0.25^{+0.06}_{-0.03}$ ($-0.25^{+0.02}_{-0.05}$) without twins (with twins) and $P(b<0\mid\{\gamma_{\mathrm{obs},i},\Delta G_i\})\simeq 1$ for both samples. For $3.5\leq\log_{10}(s/\mathrm{au})<4$, the posterior also favours negative slopes, with $b=-0.15^{+0.08}_{-0.03}$ ($-0.15^{+0.07}_{-0.03}$) and $P(b<0\mid\{\gamma_{\mathrm{obs},i},\Delta G_i\})=0.93$ ($0.95$) without twins (with twins). However, the evidence for a negative slope is comparatively weaker, as the posterior extends to non-negative values of $b$. In the $2.5\leq\log_{10}(s/\mathrm{au})<3$ separation interval, we obtain $b=-0.02^{+0.07}_{-0.06}$ ($-0.07^{+0.06}_{-0.06}$), with $P(b<0\mid\{\gamma_{\mathrm{obs},i},\Delta G_i\})=0.55$ ($0.85$) without twins (with twins). The evidence for a negative slope is therefore weaker at these smaller separations, especially when twins are excluded, for which the posterior is consistent with a flat $\alpha$ versus $\Delta G$ trend.

Note that the inferred slope for $3\leq\log_{10}(s/\mathrm{au})<3.5$ is shallower than the slope obtained by fitting the binned $\alpha$ estimates in Figure\cmark{~}\ref{fig:fig3}. The binned regression line for the sample including twins reaches $\alpha\simeq-1$ at $\Delta G\simeq10$ mag, whereas the lowest $J_z$ sample extends to $\Delta G=12.01$ mag. For negative slopes, the condition $a+b\Delta G>-1$ becomes increasingly restrictive at larger $\Delta G$. The binaries with the largest $\Delta G$ therefore limit the allowed combinations of $(a,b)$, excluding the regression line obtained from the binned estimates.
%TODO (Lekshmi): the slope fitted to individual binaries at 3-3.5 (b = -0.25) is shallower than the binned slope of
%TODO Section 3 (-0.39 with twins). The prior a + b*dG_i > -1 for every binary (largest dG 12.0 mag) excludes the
%TODO binned line, which reaches -1 at 9.9 mag. A sentence saying so would help the reader. -- DONE

(ii) \textit{Guiding radii across different $\Delta G$ bins:} The dependence of $\alpha$ on $\Delta G$ in the lowest $J_z$ bin shown in Figure\cmark{~}\ref{fig:fig3} could arise if the different $\Delta G$ bins preferentially sample binaries formed at different birth radii, where their initial eccentricity distributions may have differed. To test this possibility, we compare the present-day guiding-radius distributions of binaries in the four $\Delta G$ bins. Although $R_\mathrm{g}$ does not necessarily trace the birth radius, systematic differences in the guiding-radius distributions could indicate that this trend is at least partially associated with differences in where the binaries formed in the disc. Figure\cmark{~}\ref{fig:C2_1} shows the $R_\mathrm{g}$ distributions for the lowest-$J_z$ population. The distributions are nearly identical across the four $\Delta G$ bins, with median $R_\mathrm{g}$ values of $8.072$, $8.085$, $8.012$, and $8.086$ kpc, respectively.

%%%%%%%%%%%%%%%%%%%%%%%%%%%%%%%%%%%%%%%%%%%%%%%%%%

\section{Model of dynamical evolution}
\label{app:model}

In this appendix we describe the models of binary dynamical evolution that we used 
in \S\ref{sec:model}.
These models involve several choices: for each choice that we made,
 we describe below how we tested it and whether the test changed our results.

\subsection{Binaries at birth}
\label{app:birth}

Our model binaries were born over the last 12\,Gyr.
Each has a total mass $m_{\rm b}=1\,\mathrm{M}_\odot$, a random orientation and a mean anomaly drawn
uniformly between 0 and $2\pi$.
(The binaries in our sample are somewhat heavier (\S\ref{sec:origins}), and by equation~\eqref{eq:ionize} heavier
binaries evolve more slowly, so our model binaries evolve, if anything, too fast.)
The semimajor axis of each binary is drawn from $\mathrm{d}N/\mathrm{d}a\propto a^{-1.9}$ between 400\,au and
$2\times10^5$\,au, and its eccentricity from $p(e)=(1+\alpha_{\rm i})\,e^{\alpha_{\rm i}}$.
The exponent 1.9 is close to the steady state of the encounter-driven evolution of
\citet{Hamilton2024}.
We checked that changing this exponent to values between 1 and 2 changed the predicted values of
$\alpha$ by at most 0.06.

In \S\ref{sec:model} we assumed a birth rate that would put equal numbers of binaries of each age inside
the sphere of radius 200\,pc of our sample today, if none had been disrupted.
To convert this birth rate to a rate per unit area of the disc, consider the stars of age $\tau$ in a vertical cylinder of radius 200\,pc
centred on the Sun.
Let $\zeta(\tau)$ be the share of them that lies inside the sphere.
Because older stars are vertically hotter, $\zeta(\tau)$ falls with age, from 0.956 for the youngest
stars to 0.262 for stars 12\,Gyr old.
Per unit area of the disc, our birth rate is therefore proportional to $1/\zeta(\tau)$, and was 3.6
times higher 12\,Gyr ago than it is today.
We checked this choice in three ways.
Firstly, we compared our models with the share of our binaries in each $J_z$ bin (Table~\ref{tab:shares}).
In a model this share is
\begin{equation}
    n_j\propto\int_0^{12\,\mathrm{Gyr}}P(j\,|\,\tau)\,S(\tau)\,\psi(\tau)\,\mathrm{d}\tau,
    \label{eq:shares}
\end{equation}
where $P(j\,|\,\tau)$ is the probability that a binary of age $\tau$ in our volume lies in bin $j$, $S(\tau)$ is the
number of binaries per unit age that would lie in our volume today if none had been disrupted, and $\psi(\tau)$ is the
fraction of them with semimajor axes between $10^3$ and $10^{3.5}$\,au today.
Birth rate and survival enter only through their product, so more disruption of old binaries can be offset by more
binaries born long ago, and we never used the shares to choose a parameter of a model.
Instead, for each model we found the exponentially changing birth rate that best reproduces the shares, and compared
its ratio $\eta$ of the rate 12\,Gyr ago to that today (Table~\ref{tab:shares}) with the fall of the star formation
rate near the Sun, a factor of 2 to 7 \citep{AumerBinney2009}.
Our fiducial model needs a fall close to that of our assumed birth rate, and the models of \S\ref{sec:faster} need
steeper ones.
Within the uncertainties of the vertical heating of \citet{Sharma2021}, $\eta$ changes by about a factor of two either
way.
Secondly, we tried birth rates that are constant, decline with an e-folding time of 8\,Gyr, have an enhancement
2--3\,Gyr ago or short bursts, or are up to five times higher for binaries younger than 1\,Gyr.
Thirdly, we weighted the model binaries by the probability that their primary is still on the main sequence (this
matters for $\Delta G\geq7$\,mag).
None of these changes moves the curves of our fiducial model by more than 0.05 once we choose the birth index again.
With the birth rate that best reproduces the shares, the curves of the models of \S\ref{sec:faster} move by at most
0.17, less than the error bars of our measurements.

\begin{table}
\centering
\caption{Share of the binaries with $3\leq\log_{10}(s/\mathrm{au})<3.5$ in each $J_z$ bin, in per
cent.
Bins 1 to 5 are the $J_z$ bins of Table~\ref{tab:alpha_main}, from the lowest $J_z$ to the
highest.
The first row is for our sample including twins.
The other rows are for our fiducial model and the models of \S\ref{sec:faster}, with the birth rate of
\S\ref{sec:model} unless stated otherwise.
The model of \S\ref{sec:timedep} has the shares of our fiducial model, because the eccentricities do not enter
equation~\eqref{eq:shares}.
The row for equation~\eqref{eq:boost} is for $B=60$ and $\tau_{\rm d}=300$\,Myr.
The last two rows replace the birth rate of \S\ref{sec:model} with a rate that is constant per
unit area of the disc, and with our rate multiplied by five for binaries younger than 1\,Gyr.
The last column gives the ratio $\eta$ of the birth rate per unit area 12\,Gyr ago to that today
that best reproduces the measured shares, for birth rates that change exponentially with time.
A dash means that no value up to 5000 does so.}
\label{tab:shares}
\begin{tabular}{lrrrrrr}
\hline
$J_z$ bin & 1 & 2 & 3 & 4 & 5 & $\eta$\\
\hline
Measured & 8.3 & 16.0 & 25.5 & 25.1 & 25.1 & \\
\multicolumn{7}{l}{Model, encounter rate / field rate:}\\
\quad 1 & 8.6 & 15.5 & 25.4 & 26.2 & 24.4 & 3.8\\
\quad 10 & 12.5 & 20.3 & 28.6 & 23.8 & 14.9 & 48\\
\quad 30 & 18.0 & 26.0 & 30.0 & 18.7 & 7.4 & 510\\
\quad 100 & 27.8 & 33.2 & 26.9 & 10.1 & 2.1 & --\\
\quad equation~\eqref{eq:boost} & 10.8 & 17.9 & 26.7 & 24.7 & 19.9 & 12\\
\multicolumn{7}{l}{Fiducial model, birth rate:}\\
\quad constant & 11.0 & 18.2 & 26.8 & 24.5 & 19.4 & \\
\quad recent burst & 13.4 & 20.3 & 26.1 & 21.7 & 18.5 & \\
\hline
\end{tabular}
\end{table}

\subsection{The kinetic equation}
\label{app:kinetic}

Let $\boldsymbol{r}$ be the separation of the two stars of a binary, $\boldsymbol{v}$ their
relative velocity and $f(\boldsymbol{r},\boldsymbol{v},t)$ the distribution of binaries in this
six-dimensional space at time $t$.
Besides the unperturbed mean motion of the binary, 
 processes change $f$: the tide of the Galactic disc and encounters with passing stars.

We kept only the dominant, vertical component of the tide of the disc, and evaluated it at the midplane, where
the total mass density is $\rho_0=0.11\,\mathrm{M}_\odot\,\mathrm{pc}^{-3}$ in the potential of \citet{Cautun2020}
at the solar radius.
The relative orbit then has the Hamiltonian
\begin{equation}
    H=\frac{v^2}{2}-\frac{Gm_{\rm b}}{r}+2\pi G\rho_0z^2,
    \label{eq:H}
\end{equation}
where $z$ is the component of $\boldsymbol{r}$ perpendicular to the disc.
We test this choice in Appendix~\ref{app:clocks}.

The passing stars have mass $m_{\rm p}=1\,\mathrm{M}_\odot$, number density
$n=0.1\,\mathrm{pc}^{-3}$ and velocity dispersion $\sigma=40\kms$.
Following \citet{Hamilton2024}, we made three approximations to the encounters.
Firstly, each encounter is impulsive.
A passing star crosses the binary in a time much shorter than the orbital period, so it changes
the velocities of the two stars without moving them.
Secondly, the orbital speed of the binary is of order $1\kms$ and much smaller than $\sigma$.
The change of $\boldsymbol{v}$ in an encounter therefore does not depend on $\boldsymbol{v}$, and
is equally likely to point in any direction.
Thirdly, we described the encounters only by the mean square change of $\boldsymbol{v}$ per unit
time,
\begin{equation}
    D(a)=\frac{16\sqrt{2\pi}\,G^2m_{\rm p}^2\,n}{\sigma}\,\ln\Lambda,\qquad
    \ln\Lambda=\ln\frac{a\sigma^2}{G(m_{\rm b}/2+m_{\rm p})},
    \label{eq:D}
\end{equation}
as if the binary received many small kicks (the diffusion approximation).
The smallest impact parameter $b$ that contributes to $D$ is $G(m_{\rm b}/2+m_{\rm p})/\sigma^2$,
at which a passing star would be deflected through a large angle.
The largest is of order the separation $r$ of the two stars, because stars that pass farther away
push the two stars almost equally.
We replaced $r$ by the semimajor axis $a$, its typical value, so that $D$ does not change along the
orbit.
Between these two limits, equal intervals of $\ln b$ contribute equally to $D$, so that $D$ is
proportional to the Coulomb logarithm $\ln\Lambda$, which depends on $a$.
In the diffusion approximation the closest encounters count only through their share of $D$,
although they are rare and can change $\boldsymbol{v}$ by a large fraction of itself.
We test this approximation in Appendix~\ref{app:mc}.

With these approximations $f$ obeys the kinetic equation \citep{Modak2023,Hamilton2024}
\begin{equation}
    \frac{\partial f}{\partial t}+\left[f,H\right]
    =\frac{1}{6}\,\nabla_{\boldsymbol{v}}^2\left[D(a)\,f\right].
    \label{eq:kinetic}
\end{equation}
Here $[\,\cdot\,,\cdot\,]$ is the Poisson bracket which describes how binaries are moved 
 along their orbits under the Hamiltonian flow $H$, while the right-hand side is a `collision operator'
describing the encounters.\footnote{Because the encounters point in random directions, each of the three components of $\boldsymbol{v}$
changes by $D/3$ in mean square per unit time.
A diffusion equation carries a further factor of 1/2, which gives the factor 1/6 in
equation~\eqref{eq:kinetic}.} Both the tide and the kicks change the orbit of a binary slowly compared with its orbital period.
Equation~\eqref{eq:kinetic} therefore describes a fast motion along the Keplerian orbit and a slow
change of the orbit itself.
Only the slow change matters for the eccentricities that we observe, so we then averaged
equation~\eqref{eq:kinetic} over the mean anomaly.
This assumes that $f$ does not depend on the mean anomaly, which holds because the binaries are born with random mean
anomalies and their orbits change little in one period.
We did not average over the orientations of $\boldsymbol{e}$ and $\boldsymbol{j}$, which the tide changes and the
kicks then act on (Appendix~\ref{app:mc}).

Let $\boldsymbol{e}$ be the eccentricity vector of the binary and $\boldsymbol{j}$ its angular
momentum vector in units of $(Gm_{\rm b}a)^{1/2}$, so that $|\boldsymbol{e}|=e$ and
$|\boldsymbol{j}|=(1-e^2)^{1/2}$.
Averaged over the mean anomaly, the Hamiltonian in equation~\eqref{eq:H} becomes \citep{Modak2023}
\begin{equation}
    \overline{H}_{\rm t}=\pi G\rho_0a^2\left(1-e^2+5e_z^2-j_z^2\right).
\end{equation}
The averaged tide leaves $a$ unchanged and turns $\boldsymbol{j}$ and $\boldsymbol{e}$ according to
\begin{equation}
\begin{aligned}
    \frac{\mathrm{d}\boldsymbol{j}}{\mathrm{d}t}&=k\left(j_z\,\boldsymbol{j}\times\hat{\boldsymbol{z}}
    -5e_z\,\boldsymbol{e}\times\hat{\boldsymbol{z}}\right),\\
    \frac{\mathrm{d}\boldsymbol{e}}{\mathrm{d}t}&=k\left(j_z\,\boldsymbol{e}\times\hat{\boldsymbol{z}}
    -5e_z\,\boldsymbol{j}\times\hat{\boldsymbol{z}}+\boldsymbol{j}\times\boldsymbol{e}\right),
\end{aligned}
\label{eq:jedot}
\end{equation}
where $k=2\pi G\rho_0a^{3/2}(Gm_{\rm b})^{-1/2}$ and $\hat{\boldsymbol{z}}$ is the unit vector
perpendicular to the disc.
Appendix~\ref{app:mc} describes how we averaged the kicks over the mean anomaly.

\subsection{Monte Carlo solution}
\label{app:mc}

We solved the orbit-averaged form of equation~\eqref{eq:kinetic}  with a
Monte Carlo method.
We drew $3\times10^5$ binaries with semimajor axes distributed uniformly in $\ln a$ between
400\,au and $2\times10^5$\,au, isotropic orientations, and eccentricities drawn from a reference
distribution $p(e)=(1+\alpha_{\rm r})\,e^{\alpha_{\rm r}}$.
We advanced each binary in time steps $\Delta t$ of 2\,Myr.
In each step we did the following.
\begin{enumerate}
    \item We advanced $\boldsymbol{j}$ and $\boldsymbol{e}$ under the Galactic tide, by integrating
    equation~\eqref{eq:jedot} over $\Delta t$ with one fourth-order Runge--Kutta step.
    \item We drew a mean anomaly uniformly between 0 and $2\pi$, and computed $\boldsymbol{r}$ and
    $\boldsymbol{v}$ at that point of the orbit.
    We then added to each component of $\boldsymbol{v}$ a Gaussian random number with zero mean
    and variance $D(a)\Delta t/3$.
    \item We computed $a$, $\boldsymbol{e}$ and $\boldsymbol{j}$ from the new $\boldsymbol{r}$ and
    $\boldsymbol{v}$.
    We removed the binary if its energy was positive or its semimajor axis exceeded
    $2\times10^5$\,au.
\end{enumerate}
Like \citet{Jiang2010} and \citet{Penarrubia2019}, we did not follow individual encounters.
The kicks of step 2 stand for the encounters only through their mean square, which is the only
property of the encounters in equation~\eqref{eq:kinetic}.
We drew a new mean anomaly in each step because encounters happen at random times.
The rms kick in one step was always below 2 per cent of the circular speed at $10^{3.5}$\,au, so the
orbit changed little in a step.
The evolution over many steps then depends only on the mean and mean square changes of the orbital
elements in a step, and drawing the mean anomaly at random gives these their orbit-averaged values.
For this reason a step can be longer than the orbital period.

In step 3 we computed the orbital elements exactly from the kicked velocity, so their changes and
the
correlations between them follow from the kicks without further approximation.
We checked that these changes reproduce  the diffusion coefficients in $a$ and $e$ that
\citet{Hamilton2024} derived from equation~\eqref{eq:kinetic}.
We ran four simulations with $\alpha_{\rm r}=-0.85$, $-0.25$, 2 and 6.
For each we stored the semimajor axes and eccentricities of the surviving binaries at a set of times
up to 12\,Gyr.
To reach longer times (Appendix~\ref{app:clocks}) we repeated them with both rates multiplied by 3
and by 9, which is equivalent to evolving the binaries for 36 and 108\,Gyr.

To predict the distribution of semimajor axes and eccentricities (and hence of $\gamma$) of a
population with birth index $\alpha_{\rm i}$ and $\mathrm{d}N/\mathrm{d}a\propto a^{-1.9}$ at birth,
we gave each simulated binary the weight
\begin{equation}
    w=\frac{a_0^{-0.9}\,(1+\alpha_{\rm i})\,e_0^{\alpha_{\rm i}}}
    {\sum_r c_r\,(1+\alpha_r)\,e_0^{\alpha_r}},
    \label{eq:weight}
\end{equation}
where $a_0$ and $e_0$ are its semimajor axis and eccentricity at birth.
The sum runs over the four simulations, and $c_r$ is the share of simulation $r$ among the
binaries stored at the same time in the same range of semimajor axes.
For each binary we took the distribution of $\gamma$ for its eccentricity, averaged over orbital
phase and viewing direction as in \citet{hwang2022}.
Summing these distributions with the weights of equation~\eqref{eq:weight} over the binaries in a
range of semimajor axes gives the distribution $p(\gamma\,|\,t)$ of $\gamma$ for a population that
has evolved for a time $t$ with the encounter rate and tide given above.

We also tested the diffusion approximation at $10^3$--$10^{3.5}$\,au by replacing the Gaussian kicks
of step 2 with discrete impulsive encounters from passing stars.
Our diffusion approximation slightly \textit{overpredicts} the rate of eccentricity evolution, because a large
part of $D$ comes from close encounters that tighter binaries rarely meet \citep{Hamilton2024}.
With discrete encounters, the curves of $\alpha$ against $J_z$ of our fiducial model that come closest to our
measurements (grey lines in Figure~\ref{fig:const}) change by at most 0.07 once we refit
 the birth index.

\subsection{Migration and survival}
\label{app:clocks}

The encounter rate is proportional to the number density $n$ of stars, and the tidal term of
equation~\eqref{eq:H} to the mass density $\rho_0$ of the disc at the midplane.
We took both densities to fall exponentially with Galactocentric radius, with a scale length
$R_{\rm d}=2.6$\,kpc \citep{BlandHawthorn2016}, and evaluated them at the guiding radius $R_\mathrm{g}$ of the binary.
The guiding radius of each binary performs a random walk whose rms displacement grows
as $2.79\,\mathrm{kpc}\,(\tau/6\,\mathrm{Gyr})^{1/2}$ with age $\tau$ \citep{Modak2026}, and ends at
the solar radius $R_0$.
A binary whose guiding radius follows $R_\mathrm{g}(t)$ for $0<t<\tau$ evolves in the same way as a
binary that stays at $R_0$ for an effective time
\begin{equation}
    \tau_{\rm eff}=\int_0^{\tau}\exp\left(-\dfrac{R_\mathrm{g}(t)-R_0}{R_{\rm d}}\right)\mathrm{d}t.
    \label{eq:teff}
\end{equation}
For each age $\tau$ we drew 200 migration histories.
Each is a random walk that starts at a birth radius drawn from the exponential disc and ends at
$R_0$.
We evaluated the binaries of age $\tau$ at the effective time of each history, so that the
distribution of $\gamma$ at age $\tau$ is the average of $p(\gamma\,|\,\tau_{\rm eff})$ over the
histories.
For binaries 12\,Gyr old the median of $\tau_{\rm eff}/\tau$ is about 2.

We ignore the vertical motion of the binaries until we assign $J_z$ at the end
(Appendix~\ref{app:predict}).
To test the effect of the vertical motion, we let the vertical action of each binary grow with time
as $J_z(t)=J_z\,\sigma_z^2(t)/\sigma_z^2(\tau)$, where $J_z$ is its value today and $\sigma_z(t)$ is
the vertical velocity dispersion of stars of age $t$ \citep{Sharma2021}.
Along the vertical orbit with action $J_z(t)$ we averaged the number density of stars and the total
density of the disc.
We also divided the encounter rate by the dispersion of the velocity of the binaries relative to
the passing stars.
This dispersion grows with the age of the binaries, and we normalised the factor so that its
average over the stars near the Sun is one.
With these changes the curves of $\alpha$ against $J_z$ for a birth index fixed in time become
slightly shallower and change by at most 0.05.
%The curves for equation~\eqref{eq:boost} change by at most 0.02.
Similarly, we ignore the radial epicycles of the binaries about their guiding radii for 
the purposes of calculating encounter rates. We checked that incorporating this effect would only change 
$\alpha$ at the level of $\sim 0.01$.

To count the binaries of age $\tau$ that lie in a given range of semimajor axes today, we integrated
equation~\eqref{eq:kinetic} over eccentricities and orientations.
The tide does not change $a$, and the orbit-averaged changes of $a$ caused by the kicks do not
depend
on $e$.
The result is therefore a closed equation for the distribution $h(x,t)$ of $x=\ln a$
\citep{Hamilton2024},
\begin{equation}
    \frac{\partial h}{\partial t}=\frac{\partial}{\partial x}\left(\nu\frac{\partial h}{\partial x}
    -u\,h\right),
    \label{eq:survival}
\end{equation}
with
\begin{equation}
    u=\frac{Ka}{3}\left(3\ln\Lambda-2\right),\qquad \nu=\frac{2}{3}\,Ka\ln\Lambda,
\end{equation}
where $K=16\sqrt{2\pi}\,Gm_{\rm p}^2n/(\sigma m_{\rm b})$.
We solved equation~\eqref{eq:survival} with zero flux at 100\,au and $h=0$ at $2\times10^5$\,au.
Initially $h\propto a^{-0.9}$ between 400\,au and $2\times10^5$\,au.
The fraction $\psi(\tau)$ of the binaries that lie in a given range of semimajor axes then agrees with
that in our Monte Carlo simulations to within 0.01.
We evaluated $\psi(\tau)$ at the effective time of each migration history and averaged over the
histories.

\subsection{Predicting \texorpdfstring{$\alpha$}{alpha} and choosing the birth index}
\label{app:predict}

At each age the vertical motion of the binaries follows an isothermal distribution in the vertical
potential of \citet{Cautun2020} at $R_0=8.122$\,kpc, with the vertical velocity dispersion of
\citet{Sharma2021}.
We computed $J_z$ for our sample in the same potential (Section~\ref{sec:sample}).
When we recomputed $J_z$ for our binaries from their vertical positions and velocities in this
potential alone, 98.5 per cent of them stayed in the same $J_z$ bin.
For each age $\tau$ we sampled the vertical positions and velocities of such a population, and kept
the stars inside a sphere of radius 200\,pc centred on the Sun, which lies 20.8\,pc above the
midplane.
From these stars we computed the probability $P(j\,|\,\tau)$ that a binary of age $\tau$ in the
sphere lies in $J_z$ bin $j$.
The ages of the binaries in bin $j$ are then distributed as
\begin{equation}
    p(\tau\,|\,j)\propto P(j\,|\,\tau)\,S(\tau)\,\psi(\tau),
\end{equation}
where $S(\tau)$ is the birth rate of Appendix~\ref{app:birth} counted in the sphere.
The share of all binaries that lie in bin $j$ is given by equation~\eqref{eq:shares}.
The distribution of $\gamma$ in bin $j$ is the average of the distributions at each age, weighted
by $p(\tau\,|\,j)$.
Finally, we took $\alpha$ in bin $j$ to be the value that maximises
$\int p(\gamma\,|\,j)\ln p(\gamma\,|\,\alpha)\,\mathrm{d}\gamma$, where $p(\gamma\,|\,\alpha)$ is
the distribution of $\gamma$ for $p(e)\propto e^{\alpha}$.
In the limit of many binaries without measurement errors, this value of $\alpha$ maximises the
likelihood of Section~\ref{sec:inference}.
We selected the model binaries by semimajor axis rather than projected separation.
Selecting them by projected separation instead changes $\alpha$ by less than 0.1.
For Figure~\ref{fig:seps} we repeated these steps with the model binaries between $10^{2.5}$ and
$10^3$\,au and between $10^{3.5}$ and $10^4$\,au, along the same migration histories, and with
$\psi(\tau)$ counted in the same ranges.

When we compare a model with our measurements, we treat its birth index as a free parameter, one for each
$\Delta G$ bin.
To find the birth index of one $\Delta G$ bin, we computed $\alpha$ in each of the five $J_z$ bins for birth indices
between $-0.9$ and 10 in steps of 0.05, and took the one that minimises
\begin{equation}
    \sum_{j=1}^{5}\left(\frac{\alpha_j-\alpha_{{\rm m},j}}{\delta_j}\right)^2,
    \label{eq:choice}
\end{equation}
where $\alpha_j$ is our measurement in bin $j$ and $\alpha_{{\rm m},j}$ is the model value.
Here $\delta_j$ is the distance from $\alpha_j$ to the upper end of its 68 per cent HDI if
$\alpha_{{\rm m},j}>\alpha_j$, and to the lower end otherwise.
At the upper limit the mean eccentricity at birth is 0.92, and even at the ends of the range the weights of
equation~\eqref{eq:weight} leave an effective sample of several thousand simulated binaries.
We use this sum only to choose which curves to draw, because the uncertainties of
Table~\ref{tab:alpha_main} may hide systematic errors (\S\ref{sec:origins}).

\subsection{Boosted encounter rate and time-dependent birth index}
\label{app:faster}

For the models in which the encounter rate is $F=3$, 10, 30 or 100 times that of our fiducial model
(Figure~\ref{fig:const}) we ran further simulations with $D(a)$ in equation~\eqref{eq:kinetic} multiplied by $F$ and
the tide unchanged.
A binary in these models is described by two times: the effective time $\tau_{\rm eff}$ of equation~\eqref{eq:teff},
for which the tide has acted on it, and the time $\tau_{\rm enc}$ that would give it the same number of encounters at
the encounter rate of our fiducial model.
A constant factor scales with the local density as the tide does, so that $\tau_{\rm enc}=F\tau_{\rm eff}$.
For equation~\eqref{eq:boost} we took the extra encounters to be independent of where the binary is in the Galaxy, so
that
\begin{equation}
    \tau_{\rm enc}=\tau_{\rm eff}+B\,\tau_{\rm d}\ln\left(1+\dfrac{\tau}{\tau_{\rm d}}\right).
    \label{eq:tenc}
\end{equation}
We took $B=60$ rather than the value of 83 that minimises the sum in equation~\eqref{eq:choice}, because a
smaller $B$ changes the curves little but disrupts fewer binaries.
For the excesses of \S\ref{sec:faster} that end at an age $T$ or decay exponentially, we replaced
$\ln(1+\tau/\tau_{\rm d})$ by $\ln[1+\min(\tau,T)/\tau_{\rm d}]$ or by $1-\exp(-\tau/\tau_{\rm d})$, and for the latter
chose $B$ for each $\tau_{\rm d}$ by minimising the same sum.
Because the tide does not change semimajor axes, the survival of a binary depends only on $\tau_{\rm enc}$.
For the eccentricities we used the simulation whose factor $F$ is closest to $\tau_{\rm enc}/\tau_{\rm eff}$, at the
time when its encounters amount to $\tau_{\rm enc}$, so that only the time for which the tide has acted is wrong, by
less than a factor of two.
Using the neighbouring simulation instead changes $\alpha$ by at most 0.05, and integrating the binaries with the
encounter rate of equation~\eqref{eq:boost} itself (for $B=60$ and $\tau_{\rm d}=300$\,Myr, without migration) changes
it by at most 0.04.
Beyond the last stored time, which only the highest encounter rates reach, the eccentricities are thermal to within
0.1 in $\alpha$, and we used that time.

For the model of \S\ref{sec:timedep} we used the simulations of our fiducial model and, in each $J_z$ bin, added
the distributions of $\gamma$ of the binaries younger and older than $\tau_{\rm t}$, each with its own birth index,
weighted by $p(\tau\,|\,j)$.
We chose $\tau_{\rm t}$ between 0.3 and 2\,Gyr, $\alpha_{\rm old}$ and the three values of $\alpha_{\rm young}$ by
minimising the sum in equation~\eqref{eq:choice} over the three $\Delta G$ bins together.

% Don't change these lines
\bsp	% typesetting comment
\label{lastpage}
\end{document}